\documentclass[10pt,twocolumn]{article}

\usepackage[a4paper,margin=18mm,top=17mm,bottom=20mm]{geometry}
\usepackage[T1]{fontenc}
\usepackage[utf8]{inputenc}
\usepackage{lmodern}
\usepackage{graphicx}
\usepackage{caption}
\usepackage{subcaption}
\usepackage{amsmath,amssymb}
\usepackage{booktabs}
\usepackage{cuted}
\usepackage{xcolor}
\usepackage{stfloats}
\usepackage{hyperref}
\usepackage[super,square,sort&compress]{natbib}
\bibpunct{[}{]}{,}{s}{}{,}

\hypersetup{
  colorlinks=true,
  linkcolor=black,
  citecolor=blue!55!black,
  urlcolor=blue!55!black
}

\newcommand{\figrule}{\vspace{4pt}\hrule\vspace{8pt}}

\newenvironment{singlecolfigure}[1][tbp]
  {\begin{figure}[#1]\centering}
  {\figrule\end{figure}}

\newenvironment{doublecolfigure}[1][t]
  {\begin{figure*}[#1]\centering}
  {\figrule\end{figure*}}

\newcommand{\paperTitle}{Electrothermal control of spin-reorientation transition in Co/Fe$_3$GaTe$_2$ heterostructures}
\newcommand{\paperAuthors}{Po-Wei Chen$^{1,2,3,\dagger}$, Ming-Hsien Hsu$^{1,\dagger}$, Cheng-Ying Hsiao$^{1}$, Ming-Yang Ho$^{1}$, Masahiro Haze$^{4,*}$, Yan-Ru Chu$^{1}$, Yu-Cheng Shao$^{5}$, Po-Chun Chang$^{6,7}$, Seng-Wui Lim$^{1}$, Chen-Yu Ou$^{1}$, Ruei Chen$^{1}$, Ko-Fan Chen$^{1}$, Fang-Yuh Lo$^{1}$, Chao-Hung Du$^{6}$, Yukio Hasegawa$^{4}$, Chung-Ting Ke$^{3}$, and Wen-Chin Lin$^{1,*}$}
\newcommand{\paperAffiliations}{%
  $^{1}$Department of Physics, National Taiwan Normal University, Taipei 11677, Taiwan\\
  $^{2}$Department of Physics, The University of Osaka, Toyonaka, Osaka 560-0043, Japan\\
  $^{3}$Institute of Physics, Academia Sinica, Taipei 115201, Taiwan\\
  $^{4}$Institute for Solid State Physics, The University of Tokyo, Kashiwa, Chiba 277-8581, Japan\\
  $^{5}$National Synchrotron Radiation Research Center, Hsinchu 30076, Taiwan\\
  $^{6}$Department of Physics, Tamkang University, New Taipei City 251301, Taiwan\\
  $^{7}$Department of Materials Science and Engineering, National Yang Ming Chiao Tung University, Hsinchu 300093, Taiwan\\
  $^{\dagger}$These authors contributed equally to this work.\\
  $^{*}$Corresponding author. Email: haze@issp.u-tokyo.ac.jp (M. Haze); wclin@ntnu.edu.tw (W.-C. Lin)
}
\newcommand{\paperKeywords}{Spin-reorientation transition; Fe$_3$GaTe$_2$; van der Waals magnet; Scanning tunneling microscopy; Co heterostructure; Magnetic anisotropy; Magneto-optical Kerr effect}

\newcommand{\paperAbstract}{%
Thermally driven competition between magnetic anisotropy energies (MAEs) provides a pathway toward electrical control of van der Waals magnets. Here we demonstrate reversible thermal and electrothermal control of a spin-reorientation transition (SRT) in Co/Fe$_3$GaTe$_2$ (FGaT) heterostructures. \textit{In situ} scanning tunneling microscopy reveals a dose-dependent evolution of Co morphology on freshly cleaved FGaT, from isolated Co islands at low deposition doses to more extensive coverage featuring stripe-like structures at higher doses. The stripes are approximately parallel to Te lattice directions, suggesting substrate influence on Co growth. Temperature-dependent magneto-optical Kerr measurements show a change from out-of-plane (OOP)- to in-plane (IP)-dominated MAE. The OOP remanence decreases sharply near 300~K, where the nucleation fields change sign, while the IP loops become increasingly square upon further heating. Electrical driving induces a similar evolution of the magnetic hysteresis loops in an $\mathrm{AlO}_{x}$/Pt/Pd/Co/FGaT device, attributed to Joule heating possibly from conducting filaments in the oxide layer. The magnetic responses recover during temperature and power cycling, supporting reversible SRT. These observations are consistent with competition between the perpendicular magnetic anisotropy of FGaT and the IP magnetization preference of Co. Furthermore, polar Kerr imaging reveals that increasing electrical power facilitates magnetization reversal in FGaT at reduced reversal fields, enabling electrothermal control of FGaT devices; at a fixed assisting field of $-700$~Oe, abrupt switching occurs between 0.4 and 0.6~$\mathrm{nW}/\mathrm{\mu}\mathrm{m}^{2}$. These results support electrothermal approaches to SRT control and field-assisted magnetic switching in FGaT-based devices.
}

\newlength{\abstractBlockHeight}
\bibpunct{[}{]}{,}{s}{}{,}


\begin{document}

\begin{strip}
  \noindent
  \begin{minipage}[t]{\textwidth}
    \raggedright
    {\fontsize{17}{21}\selectfont\bfseries \paperTitle\par}
    \vspace{7pt}
    {\normalsize \paperAuthors\par}
    \vspace{5pt}
    {\footnotesize \paperAffiliations\par}
  \end{minipage}

  \vspace{12pt}
\noindent
\begin{minipage}[t]{\textwidth}
  {\bfseries Abstract\par}
  \vspace{3pt}
  {\small \paperAbstract\par}
  \vspace{6pt}
  {\footnotesize\textbf{Keywords:} \paperKeywords\par}
\end{minipage}
  \vspace{8pt}
  \hrule
  \vspace{12pt}
\end{strip}

\section{Introduction}
Electrical control of magnetism is a central goal in spintronics because it offers a route to reconfigurable magnetic states without relying solely on large write currents. In conventional magnetic heterostructures, electric fields can tune carrier density, interfacial spin-orbit coupling and magnetic anisotropy~\cite{1,2}. Heating can transiently reduce coercivity near a magnetic transition~\cite{3,4}. The emergence of two-dimensional van der Waals (vdW) magnets has expanded this idea into atomically thin systems, where clean interfaces, reduced dimensionality and strong anisotropy allow magnetic order to be coupled efficiently to gate fields, strain and interfacial exchange~\cite{5,6}.

Among these materials, Fe$_3$GaTe$_2$ (FGaT) has recently emerged as a particularly attractive room-temperature ferromagnetic metal. Bulk crystals, exfoliated flakes and wafer-scale films of FGaT exhibit a Curie temperature ($T_{\mathrm{C}}$) above room temperature, large perpendicular magnetic anisotropy (PMA), and square out-of-plane (OOP) hysteresis loops, making it one of the few vdW magnets compatible with ambient spintronic operation~\cite{7,8,9}. Moreover, its magnetic ground state is not rigid. Magnetic force microscopy and Lorentz transmission electron microscopy have revealed thickness- and field-dependent stripe domains, bubble domains and skyrmion-like spin textures in FGaT~\cite{10,11}, while non-stoichiometry and local inversion-symmetry breaking can generate Dzyaloshinskii-Moriya interaction (DMI) and stabilize chiral textures~\cite{12,13}. These findings indicate that FGaT is both a robust ferromagnet and a highly tunable magnetic host.

Interface engineering provides an additional degree of freedom for controlling FGaT. Au capping has been shown to modulate bubble and stripe domains at room temperature~\cite{14}, while Pt-, WTe$_2$- and Ti-based heterostructures enable current-induced, spin-orbit or orbital-torque switching of FGaT~\cite{15,16,17}. In parallel, conductive atomic force microscopy (cAFM) has demonstrated localized electrothermal writing of bubble domains in FGaT~\cite{18}. This approach uses a conductive probe to deliver the electrical input locally. Ferroelectric and electric-field-coupled FGaT devices further demonstrate that its $T_{\mathrm{C}}$, coercivity and PMA can be strongly modulated by interfacial polarization, strain or carrier redistribution~\cite{19,20}. Nevertheless, how a conventional metallic ferromagnet such as Co competes with FGaT across a direct magnetic interface remains comparatively unexplored. A Co/FGaT bilayer allows these competing magnetic contributions to be examined. The relative influence of Co should increase as the FGaT magnetization weakens near its $T_{\mathrm{C}}$.

\begin{doublecolfigure}
  \includegraphics[width=\textwidth]{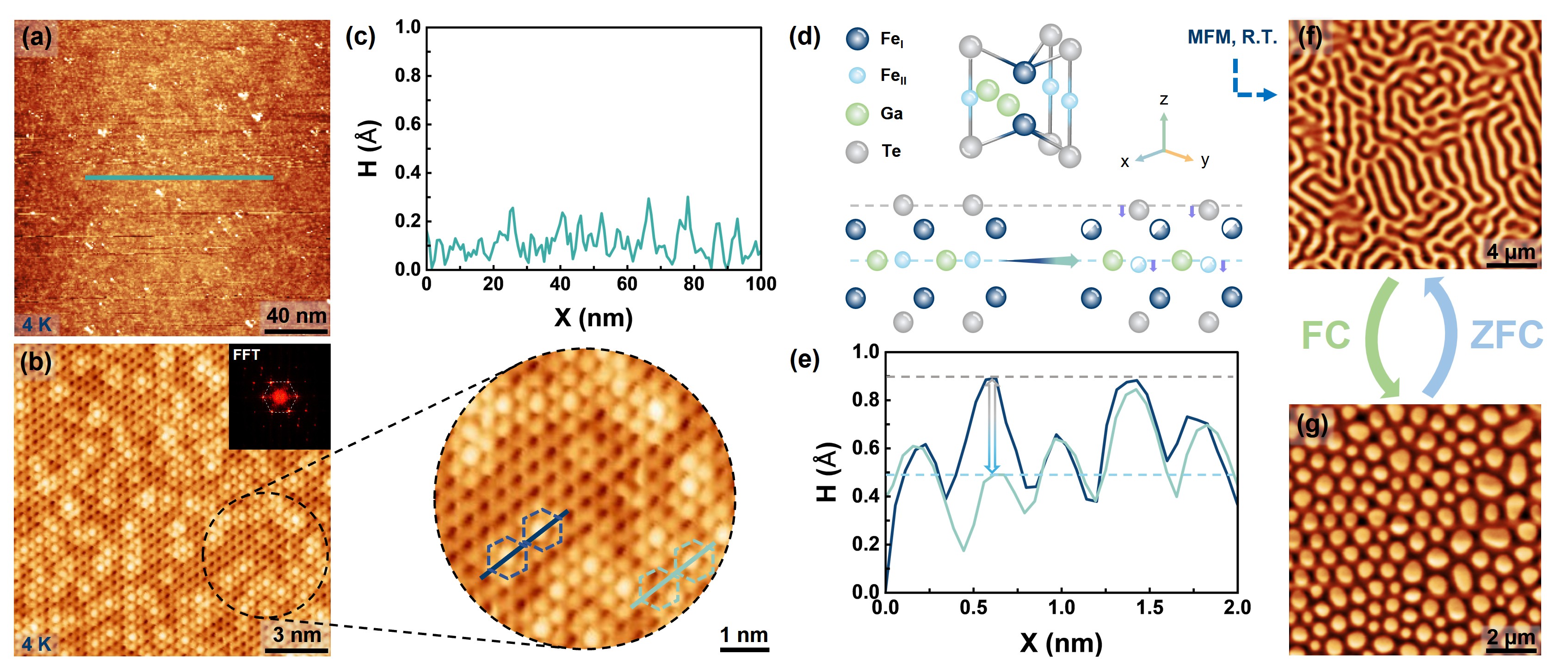}
  \caption{Surface and magnetic-domain features of cleaved Fe$_3$GaTe$_2$ (FGaT) characterized by scanning tunneling microscopy (STM) and magnetic force microscopy (MFM). (a) Large-area STM topograph acquired at 4~K with a sample bias of $+1.6$~V and a tunneling current of 200~pA. (b) Atomic-resolution STM image acquired at 4~K with a sample bias of $+0.1$~V and a tunneling current of 1000~pA, showing the hexagonal Te lattice. The fast Fourier transform (FFT) inset shows sixfold in-plane (IP) symmetry. The black dashed circle marks the region shown in the central circular enlargement. Dark- and light-blue dashed hexagons in the enlarged view mark local motifs used for line-profile comparison. (c) Height profile along the green line in (a). (d) Schematic illustration of a possible subsurface Fe-deficiency/off-centering configuration. (e) Height profiles along the dark- and light-blue lines in the central circular enlargement, plotted from left to right along the marked lines. The arrow indicates the apparent-height contrast between the central Te sites. (f,g) MFM images acquired at room temperature without an applied magnetic field after (f) zero-field cooling (ZFC) and (g) field cooling (FC) from $T>T_{\mathrm{C}}$, showing labyrinthine stripe domains and bubble domains, respectively. A field of 440~Oe was applied along the $c$ axis only during FC.}
  \label{stm_deficiency}
\end{doublecolfigure}

This competition is naturally framed in terms of a spin-reorientation transition (SRT). In ultrathin magnetic films, the easy axis is determined by a balance among magnetocrystalline anisotropy, shape anisotropy, magnetoelastic contributions and surface or interfacial anisotropy. Changing thickness, temperature or interface chemistry can therefore rotate the preferred magnetization direction~\cite{21,22,23}. In Co/NiO/Fe, temperature-driven in-plane spin reorientation in Fe can be transferred to the Co overlayer through NiO-mediated exchange~\cite{24}. In Co/FGaT, the analogous control parameter is the temperature-dependent magnetic contribution of FGaT. Heating FGaT toward its $T_{\mathrm{C}}$ is expected to weaken this contribution and alter the balance with the Co layer. The reported $T_{\mathrm{C}}$ of bulk and exfoliated FGaT lies at approximately 350--380~K~\cite{7}. This temperature range motivates using Joule heating to tune magnetic anisotropy competition near room temperature.

Here we investigate how external heating and Joule heating modify the competition between FGaT PMA and Co in-plane (IP) magnetization in Co/FGaT heterostructures. Scanning tunneling microscopy (STM) characterizes atomic-scale surface contrast and Co growth morphology, alongside magnetic force microscopy (MFM) imaging of FGaT domain textures. Temperature-dependent and electrically driven magneto-optical Kerr effect (MOKE) measurements track the evolution and reversibility of the OOP and IP magnetic responses. The correspondence between these responses supports an electrothermal interpretation of the electrically driven SRT. In a Co-free FGaT device, polar Kerr microscopy examines power-dependent, field-assisted magnetization reversal. These measurements complement cAFM-based local writing in FGaT~\cite{18} and relate the device response to heat-assisted magnetic recording, in which heating facilitates field-driven reversal~\cite{3,4}.

\section{Results}
Fig.~1 shows the surface morphology and atomic-scale features of FGaT observed by low-temperature STM. To minimize surface oxidation and contamination, the FGaT crystal was mechanically cleaved in the preparation chamber at room temperature. Large-area and atomically resolved STM images were then acquired at 4~K (Figs.~1(a) and 1(b)). The height profile along the green line in Fig.~1(a) is shown in Fig.~1(c). The large-area image and its height profile show an extended terrace without visible step edges or step-like height changes, indicating a flat vdW surface within the scanned region.

\begin{doublecolfigure}
  \includegraphics[width=\textwidth]{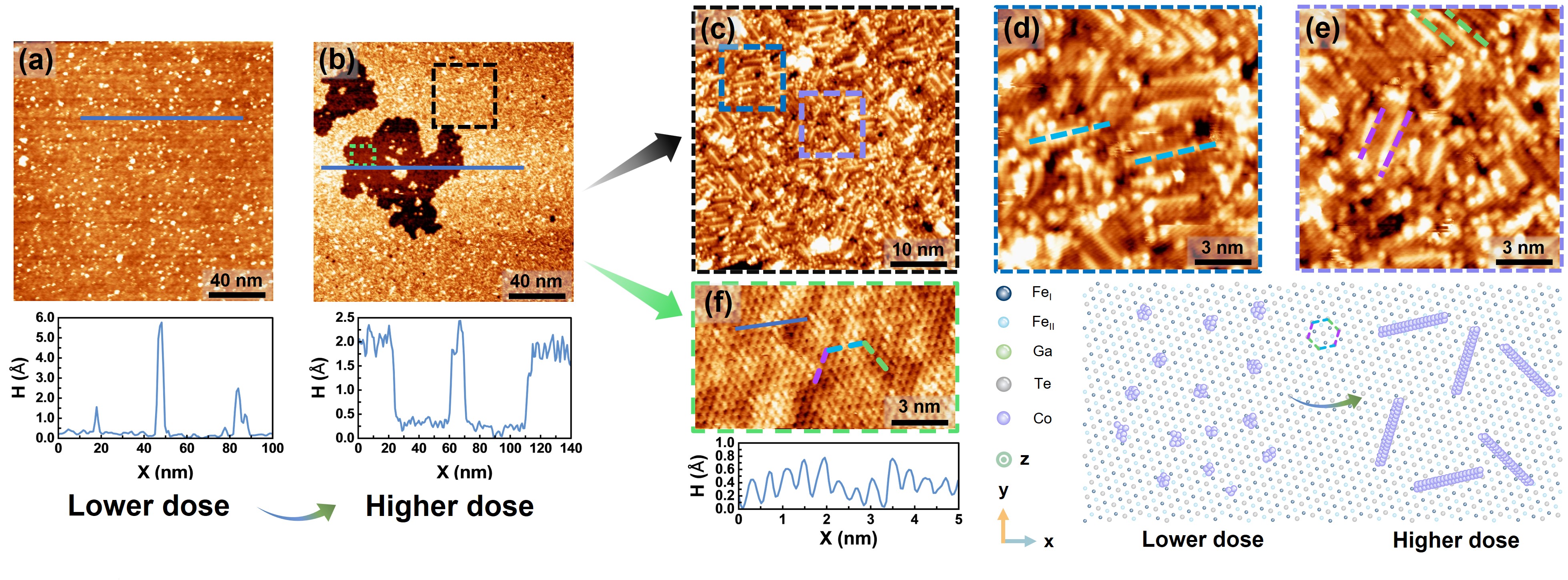}
  \caption{\textit{In situ} STM characterization of Co growth on cleaved FGaT. All STM images were acquired at 4~K. (a) Large-area STM topograph after low-dose Co deposition, acquired with a sample bias of 1.6~V and a tunneling current of 200~pA. The height profile along the blue line is shown below. (b) STM topograph after re-cleaving FGaT and depositing a higher dose of Co, acquired with a sample bias of 1.0~V and a tunneling current of 200~pA. The height profile along the blue line is shown below. (c) Enlarged STM image of the Co-covered region marked by the black dashed box in (b). The sample bias was 1.0~V and the tunneling current was 200~pA. (d,e) Enlarged STM images of the regions marked by the dark-blue and purple dashed boxes in (c), respectively. Both images were acquired with a sample bias of 1.0~V and a tunneling current of 200~pA. The light-blue, purple and green dashed lines indicate three orientations of the stripe-like Co clusters. (f) (left) Enlarged STM image of the uncovered FGaT region marked by the green dashed box in (b). The sample bias was 0.1~V and the tunneling current was 1000~pA. The height profile along the dark-blue line is shown below. The purple, light-blue and green dashed lines mark three Te atomic-row directions at the FGaT surface, corresponding to the Co stripe orientations indicated in (d,e). (right) Schematic illustration of Co growth on FGaT at different deposition doses.}
  \label{fig:co-on-fgt-stm}
\end{doublecolfigure}

The enlarged view of Fig.~1(b) resolves a periodic hexagonal lattice consistent with the expected Te-terminated surface of FGaT. The corresponding fast Fourier transform (FFT) shows sixfold in-plane symmetry, consistent with the reported hexagonal structure of FGaT \cite{7}. The dark-blue and light-blue dashed outlines each mark a pair of adjacent hexagonal motifs with different apparent-height contrasts at their central Te sites. Line profiles across these two pairs are shown in Fig.~1(e), plotted from left to right along the marked lines. The two central sites along the dark-blue profile have similar apparent heights, whereas the light-blue profile shows a lower apparent height at the left central site than at the right one.

STM apparent height depends on both surface geometry and the local electronic density of states \cite{25}. Li et al. reported that Fe deficiency in nonstoichiometric $\mathrm{Fe}_{3-x}\mathrm{GaTe}_{2}$ can induce $\mathrm{Fe}_{\mathrm{II}}$ off-centering, break inversion symmetry, and generate DMI that stabilizes N\'{e}el-type skyrmions \cite{12}. In $\mathrm{Fe}_{5}\mathrm{GeTe}_{2}$, STM and scanning tunneling spectroscopy have also shown that $\mathrm{Fe}_{\mathrm{I}}$ vacancies and split-site configurations affect the local topographic and electronic contrast of the Te-terminated surface \cite{26}. The contrast in Figs.~1(b) and 1(e) may similarly reflect local electronic or structural perturbations at the surface Te sites. A possible origin is subsurface Fe deficiency accompanied by $\mathrm{Fe}_{\mathrm{II}}$ off-centering \cite{12}, as illustrated schematically in Fig.~1(d).

Room-temperature magnetic force microscopy (MFM) images show labyrinthine stripe domains in FGaT after zero-field cooling (ZFC) (Fig.~1(f)), whereas bubble domains appear after field cooling (FC) (Fig.~1(g)). This cooling-history dependence is consistent with previous reports of bubble and skyrmion textures formed by field cooling in FGaT and $\mathrm{Fe}_{3-x}\mathrm{GaTe}_{2}$ \cite{11,12,13}.

To examine the initial growth morphology of Co on cleaved FGaT, Co was deposited \textit{in situ} in the STM preparation chamber at room temperature. The sample was then cooled to 4~K for STM imaging. The apparent Co coverage was estimated from leveled STM topographs using grain analysis in Gwyddion \cite{27} (Supplementary Fig. S1 and Table S1). After low-dose Co deposition, the surface exhibits discrete nanoscale protrusions with an apparent Co coverage of approximately 13.6\% (Fig.~2(a)). This morphology indicates the formation of isolated islands during the initial deposition stage. It is consistent with Volmer--Weber-type growth, in which three-dimensional islands form under incomplete wetting \cite{28}. Three-dimensional Co clusters attributed to Volmer--Weber growth have also been reported on pristine highly oriented pyrolytic graphite \cite{29}.

The island morphology may be related to interfacial interactions and geometrical matching between Co and the Te-terminated FGaT surface. The hexagonal close-packed (hcp) structure is the stable bulk phase of Co\cite{30}. We therefore used bulk hcp Co(0001) as a close-packed reference for a literature-based comparison. The reported IP lattice constants are approximately 2.5~\AA{} for hcp Co \cite{31} and 3.986~\AA{} for bulk FGaT \cite{7}. A one-to-one comparison of these literature values gives a geometrical mismatch of approximately 37\%, referenced to the FGaT lattice constant. Although rotated or higher-order coincidence lattices may offer alternative matching arrangements \cite{32}, this large geometrical mismatch disfavors coherent growth under the assumption of one-to-one registry and may influence Co growth on FGaT.

\begin{doublecolfigure}
  \includegraphics[width=\textwidth]{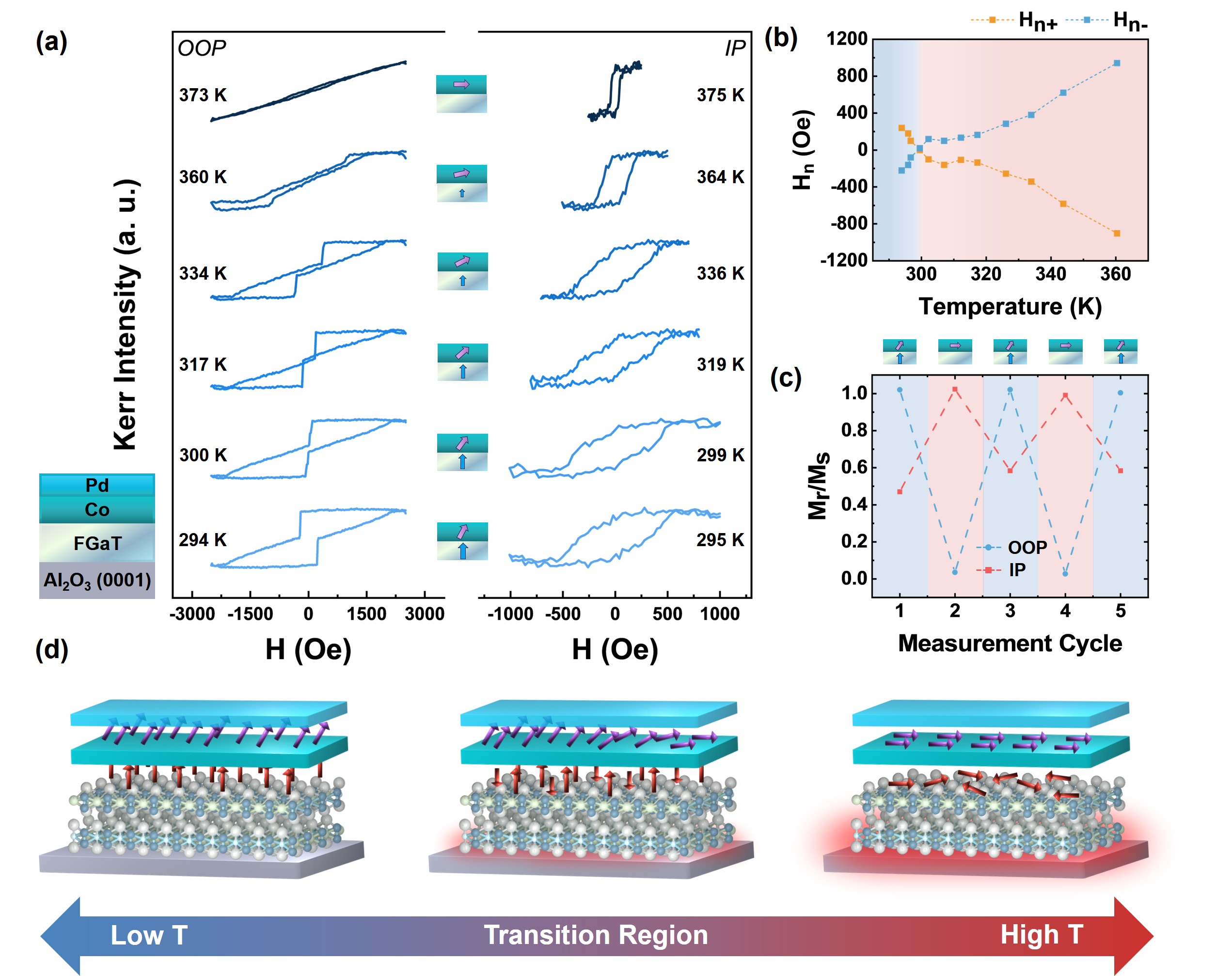}
  \caption{Temperature-driven spin-reorientation transition (SRT) in Co/FGaT. (a) Temperature-dependent out-of-plane (OOP) and in-plane (IP) MOKE hysteresis loops of the Pd(3~nm)/Co(10~nm)/FGaT(198~nm) heterostructure on an $\mathrm{Al}_{2}\mathrm{O}_{3}$(0001) substrate. (b) Temperature dependence of the nucleation fields $H_{\mathrm{n}+}$ and $H_{\mathrm{n}-}$, determined from peaks in the field derivative of the OOP Kerr signal. $H_{\mathrm{n}+}$ and $H_{\mathrm{n}-}$ correspond to the ascending and descending field-sweep branches, respectively. (c) Remanence ratios $M_{\mathrm{r}}/M_{\mathrm{s}}$ obtained from the OOP and IP MOKE loops during alternating low- and high-temperature measurements. Blue and red backgrounds indicate low- and high-temperature measurements, respectively. (d) Schematic illustration of the temperature-driven SRT.}
  \label{temp_drive}
\end{doublecolfigure}

To compare the morphology at a higher deposition dose, the crystal was returned to room temperature and re-cleaved in the preparation chamber to expose a fresh surface. A higher dose of Co was then deposited \textit{in situ}, followed by cooling to 4~K for STM imaging. The resulting surface shows more extensive Co-associated bright regions, with an apparent coverage of approximately 73.3\% (Fig.~2(b)). A higher-resolution image of the covered region marked by the black dashed box in Fig.~2(b) reveals nanoscale clusters together with stripe-like features (Fig.~2(c)).

Figs.~2(d) and 2(e) show enlarged views of the dark-blue and purple dashed boxes in Fig.~2(c), respectively. The light-blue, purple and green dashed lines indicate the orientations of the stripe-like Co clusters. The left panel of Fig.~2(f) shows an enlarged view of the uncovered region marked by the green dashed box in Fig.~2(b). This image resolves the surface Te lattice of bare FGaT, and the corresponding height profile shows atomic-scale corrugation. The purple, light-blue and green dashed lines mark three Te atomic-row directions at the FGaT surface that are approximately parallel to the Co stripe orientations indicated in Figs.~2(d) and 2(e). This orientational correspondence suggests that the arrangement of surface Te atoms may influence Co stripe growth directions at the higher deposition dose, as illustrated in the schematic on the right of Fig.~2(f).

To further investigate the magnetic response of Co/FGaT, FGaT was mechanically exfoliated onto an Al$_2$O$_3$(0001) substrate, and approximately 10~nm of Co was deposited on a flake approximately 198~nm thick (Supplementary Fig.~S2(a)) to form a magnetic bilayer. A 3~nm Pd capping layer was then deposited to reduce oxidation, giving a Pd(3~nm)/Co(10~nm)/FGaT(198~nm)/Al$_2$O$_3$(0001) structure. Temperature-dependent MOKE hysteresis loops were measured in the IP and OOP geometries. Co thin films generally favor IP magnetization unless perpendicular interfacial anisotropy is sufficiently strong to stabilize PMA~\cite{21,33}. Consistent with this preference, the Co layer in the device of Fig.~5 exhibits IP hysteresis where it extends beyond the FGaT flake (Supplementary Fig.~S3). The $T_{\mathrm C}$ of Co thin films also exceeds room temperature and that of FGaT~\cite{34}. In contrast, FGaT exhibits strong PMA, with reported flake $T_{\mathrm C}$ values of approximately 350--380~K~\cite{7}. These different preferred magnetization directions provide a basis for interpreting the Co/FGaT response in terms of competing anisotropies. A reduction of the perpendicular FGaT contribution on heating can allow the Co-associated IP response to become dominant, consistent with theoretical descriptions of temperature-induced SRT driven by competing anisotropies in thin films~\cite{35}.

Fig.~3(a) shows representative OOP and IP MOKE hysteresis loops of the heterostructure at similar temperatures; the complete temperature series for both geometries are shown in Supplementary Fig.~S4. At 294~K, the OOP loop has a large remanent Kerr signal and abrupt switching, whereas the IP loop at 295~K has a lower remanence ratio, consistent with an FGaT-dominated PMA state. With increasing temperature, the OOP remanence decreases while the IP loop becomes more square. At the highest measurement temperatures, the linear OOP response near 373 K is consistent with the vanishing magnetic contribution from FGaT as it approaches its Curie temperature, in agreement with previous reports~\cite{7}. The absence of an observable OOP hysteresis loop above this temperature indicates that the Co layer no longer exhibits perpendicular magnetic behavior. In contrast, the IP loop approaches 100\% remanence at 375 K, confirming the stable in-plane magnetization of the Co layer. These complementary evolution further support the temperature-induced SRT from the perpendicular to the IP direction~\cite{45,46}. Note that the different optical geometries and magneto-optical coefficients result in a substantially stronger polar MOKE signal, typically 5–10 times larger than the longitudinal MOKE signal, accounting for the relatively noisy IP loops. Nevertheless, these minor background variations and signal fluctuations do not affect the observed trends or the conclusion of temperature-induced magnetic easy-axis reorientation.

The reversal features in the OOP loops also shift systematically with temperature. We used the characteristic peaks in the field derivative of the Kerr signal, $\mathrm{d}I_{\mathrm K}/\mathrm{d}H$, on each sweep branch to define the nucleation fields $H_{\mathrm n}$ (Supplementary Fig.~S4(b)). The fields on the ascending branch, swept from negative to positive field, and the descending branch, swept from positive to negative field, are denoted $H_{\mathrm{n}+}$ and $H_{\mathrm{n}-}$, respectively. As shown in Fig.~3(b), both fields initially approach zero on heating and change sign at approximately 300~K. Beyond this crossing, the reversal-related Kerr features occur before the applied field changes polarity, substantially reducing the zero-field remanence. Reversed-domain nucleation followed by domain-wall expansion is a common reversal process in PMA films~\cite{36}. These processes affect the zero-field Kerr signal, so the temperature dependence of the remanence ratio reflects both magnetic anisotropy and domain reversal. The OOP loops lose squareness and the nucleation fields cross at higher temperatures in Co-free FGaT (Supplementary Fig.~S5) than in Co/FGaT, consistent with a contribution from Co to magnetic reversal. MFM images show domains after a sweep to a reversal field near $H_{\mathrm n}$ and a return to zero along a first-order reversal curve (FORC)~\cite{47}, whereas no domains were resolved after a direct return from saturation (Supplementary Fig.~S6). The nucleation fields and remanent domain patterns also vary with flake thickness (Supplementary Fig.~S7).

Fig.~3(d) summarizes this interpretation schematically in terms of competing anisotropies: the FGaT PMA dominates at low temperature ($<300$~K), OOP and IP contributions coexist in the transition region, and the Co-associated IP contribution dominates at high temperature ($>370$~K).

\begin{singlecolfigure}[!t]
  \includegraphics[width=\columnwidth]{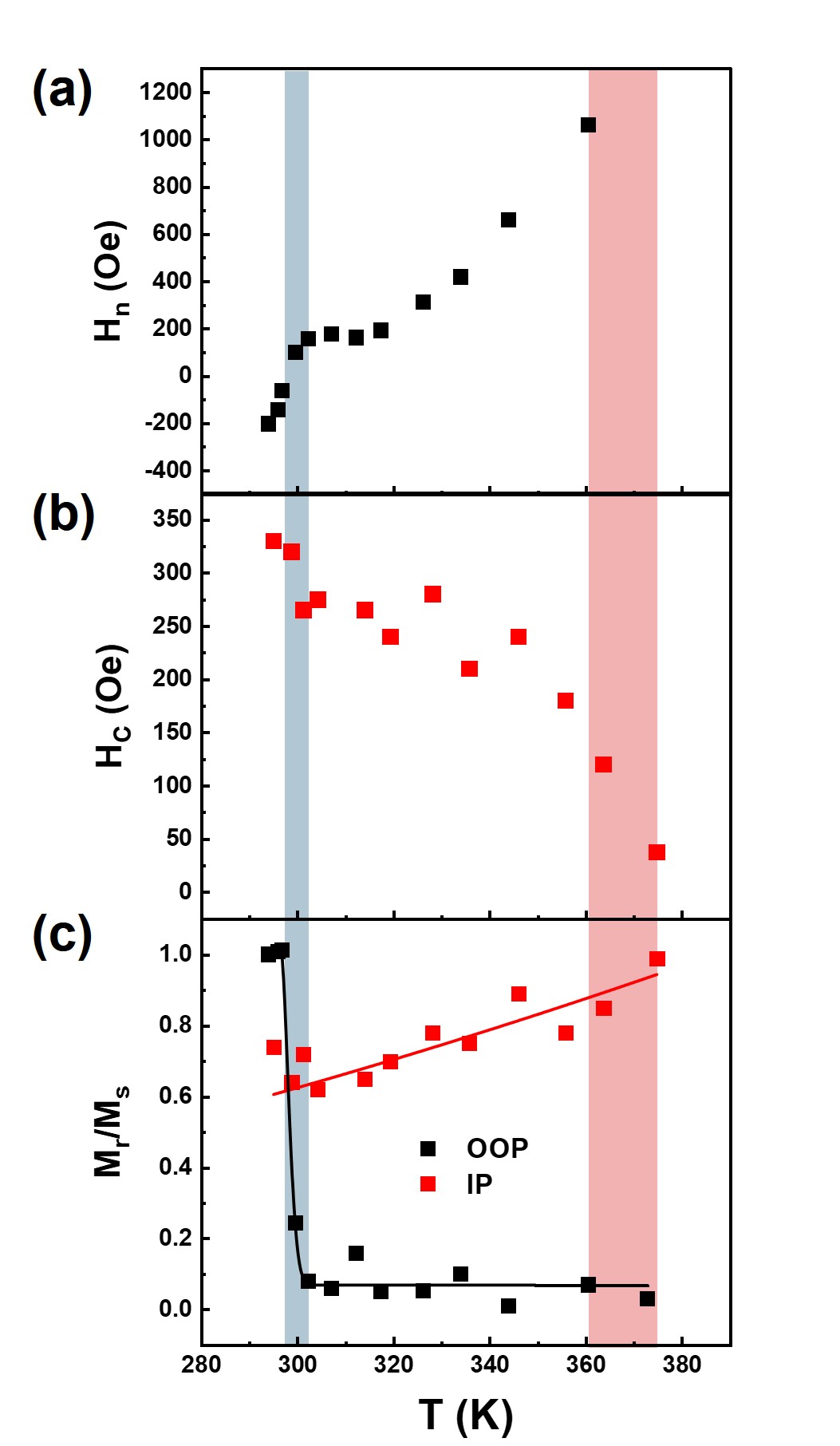}
  \caption{Temperature dependence of the nucleation field, coercivity and remanence ratio in the Co/FGaT heterostructure. (a) Nucleation field $H_{\mathrm{n}-}$ determined from the descending field-sweep branches of the out-of-plane (OOP) MOKE loops. (b) Coercivity $H_{\mathrm{c}}$ obtained from the in-plane (IP) MOKE loops. (c) Remanence ratios $M_{\mathrm{r}}/M_{\mathrm{s}}$ obtained from the OOP (black) and IP (red) MOKE loops.}
  \label{temperature_dependent_analysis}
\end{singlecolfigure}

To examine the recovery of the low- and high-temperature magnetic responses, we compared the remanence ratio $M_{\mathrm{r}}/M_{\mathrm{s}}$ at these two endpoints (Fig.~3(c)), calculated as the zero-field Kerr signal divided by the Kerr signal at the saturation field. In the low-temperature OOP-dominated state, the OOP $M_{\mathrm{r}}/M_{\mathrm{s}}$ is close to 1.0, whereas the IP ratio remains approximately 0.5. This finite IP remanence may include an IP contribution from regions of the Co layer away from the Co/FGaT interface. After heating, the IP $M_{\mathrm{r}}/M_{\mathrm{s}}$ approaches 1.0 and the OOP ratio decreases to near zero. The recovery of both endpoint remanence responses over five alternating low- and high-temperature measurements supports a reversible thermally driven SRT in the Co/FGaT heterostructure within the measured temperature range.

\begin{doublecolfigure}
  \includegraphics[width=\textwidth]{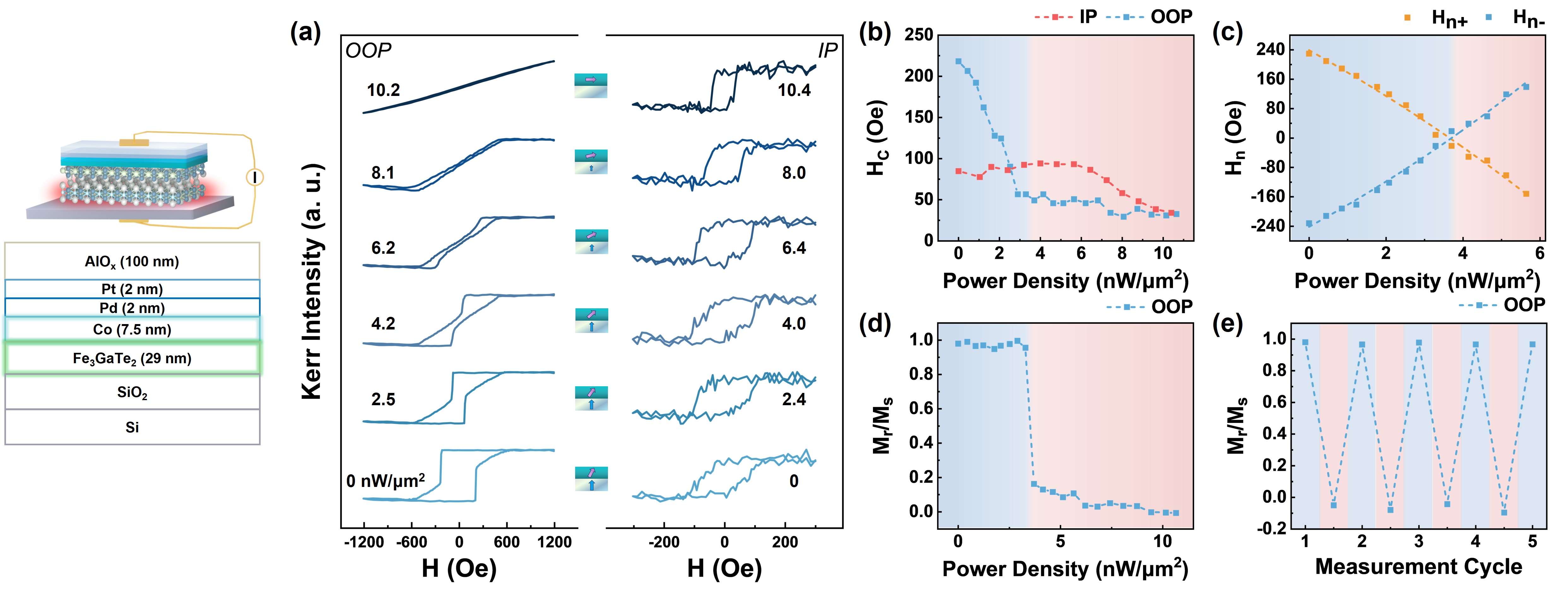}
  \caption{Electrothermal control of the spin-reorientation transition in a Co/FGaT device. Left: schematic of the $\mathrm{AlO}_{x}$(100~nm)/Pt(2~nm)/Pd(2~nm)/Co(7.5~nm)/FGaT(29~nm)/$\mathrm{SiO}_{2}$/Si multilayer device. (a) MOKE hysteresis loops measured in the out-of-plane (OOP) and in-plane (IP) geometries at varying power densities. (b) Coercivity obtained from the OOP and IP MOKE loops as a function of power density. (c) Nucleation fields $H_{\mathrm{n}+}$ and $H_{\mathrm{n}-}$, determined from peaks in the field derivative of the OOP Kerr signal, as a function of power density. $H_{\mathrm{n}+}$ and $H_{\mathrm{n}-}$ correspond to the ascending and descending field-sweep branches, respectively. (d) OOP remanence ratio $M_{\mathrm{r}}/M_{\mathrm{s}}$ as a function of power density. (e) Repeated power-cycling measurements of the OOP $M_{\mathrm{r}}/M_{\mathrm{s}}$.}
  \label{electro_control}
\end{doublecolfigure}

To illustrate the temperature evolution of the competing magnetic responses, Fig.~4 compares the nucleation field $H_{\mathrm{n}-}$ (Fig.~4(a)), IP coercivity $H_{\mathrm{c}}$ (Fig.~4(b)), and OOP and IP remanence ratios $M_{\mathrm{r}}/M_{\mathrm{s}}$ (Fig.~4(c)). In the nucleation-field crossing region near 300~K, the OOP remanence ratio decreases sharply from approximately 1.0, while the IP $H_{\mathrm{c}}$ also decreases abruptly. Above this region, $H_{\mathrm{n}-}$ generally shifts toward larger positive values on heating, whereas the IP $H_{\mathrm{c}}$ shows an overall decrease. The relatively thick FGaT flake has strong PMA, and its diminishing magnetic contribution on heating may affect the overall magnetization reversal of the Co/FGaT heterostructure and thereby reduce the IP $H_{\mathrm{c}}$. After the nucleation fields cross, the reversal features of the OOP loops make the OOP $H_{\mathrm{c}}$ difficult to quantify precisely. We therefore present only the IP $H_{\mathrm{c}}$ here. After the sign change, the reversal features appear before the applied field reaches zero, reducing the OOP zero-field remanence and keeping its ratio low. By contrast, the IP remanence ratio increases overall on heating and eventually approaches 1.0. These changes are consistent with the SRT from an OOP-dominated to an IP-dominated response shown in Fig.~3(a).

To extend the temperature-driven SRT to an electrically controlled device, we fabricated an AlO$_x$(100~nm)/Pt(2~nm)/Pd(2~nm)/Co(7.5~nm)/\allowbreak{FGaT(29~nm)/SiO$_2$/Si} multilayer (Fig.~5, left). Compared with the sample in Fig.~3, this device uses a thinner FGaT flake (see the AFM image in Supplementary Fig.~S2) and a slightly thinner Co layer. The Pt/Pd bilayer provides metallic protection between Co/FGaT and the oxide deposition environment. The Pd/Co interface may contribute to PMA, but the relative influence of interfacial anisotropy decreases with increasing Co thickness compared with the shape anisotropy favoring IP magnetization~\cite{21,33}. The AlO$_x$ layer was intended to electrically isolate the magnetic stack, but I--V measurements of a thickness-matched Co-free control device with an AlO$_x$/Pt top layer prepared under similar conditions show measurable current (Supplementary Fig.~S8(a)) and high-resistance hysteretic behavior (inset of Supplementary Fig.~S8(a)). This response is consistent with reported resistive-switching behavior in oxides~\cite{37,38}; conduction through filaments formed during electroforming is a possible mechanism~\cite{37}. In addition, the nucleation field and coercivity in this control device show similar trends under positive and negative bias and decrease as the dissipated power increases (Supplementary Fig.~S8(b--d)), consistent with the similar loop evolution under both bias polarities in Supplementary Fig.~S9(a,e). These results support a Joule-heating interpretation of the magnetic response to electrical input, which we therefore treat as an electrothermal drive. This interpretation is also consistent with heat-assisted magnetic writing, in which heating reduces coercivity and facilitates field-assisted reversal~\cite{3,4}. During the measurements, the electrical input was applied in a two-terminal configuration through two silver-paste contacts, one on the top AlO$_x$ surface and the other on the abraded back of the SiO$_2$/Si substrate. Silver paste was used instead of wire bonding to avoid puncturing the thin oxide layer. In the Co/FGaT overlap region, the path from the top contact to the back contact passes through AlO$_x$/Pt/Pd/Co/FGaT and the underlying SiO$_2$/Si substrate. Current flow through the device generates Joule heat. We use the areal power density as the control parameter, defined as the dissipated power $P=IV$ divided by the substrate area (5~mm $\times$ 5~mm).

Fig.~5(a) shows representative power-dependent MOKE hysteresis loops measured in the OOP and IP geometries. At $0~\mathrm{nW}/\mathrm{\mu}\mathrm{m}^{2}$, the OOP loop has a large remanent Kerr signal and an abrupt switching feature, indicating an FGaT-dominated PMA response. As the power density increases, the OOP loop becomes less square and the remanent Kerr signal is suppressed, approaching a linear, hard-axis-like response at $10.2~\mathrm{nW}/\mathrm{\mu}\mathrm{m}^{2}$. The IP loops evolve in a complementary manner, from low-remanence loops toward nearly square, easy-axis-like loops. As in Fig.~3, finite IP coercivity is already present at zero power, consistent with a residual IP component in the heterostructure that may include contributions from Co away from the Co/FGaT interface. The overall evolution resembles that in the temperature-dependent measurements of Fig.~3, supporting an interpretation in which electrical input and external heating alter the magnetic response through similar anisotropy competition: the FGaT PMA contribution weakens relative to the Co-dominated IP contribution \cite{7,35}.

Fig.~5(b--d) further quantifies the OOP-to-IP crossover. The OOP coercivity decreases rapidly at low power, whereas the IP coercivity remains nearly unchanged over the lower-power range and begins to decrease appreciably around 6.4~nW/$\mathrm{\mu}$m$^2$ (Fig.~5(b)). The different power ranges over which the coercivities decrease may reflect a residual influence of FGaT PMA on IP reversal that weakens with further Joule heating. This decrease in IP coercivity is similar to that observed in the temperature-dependent measurements in Fig.~4(b). The nucleation fields $H_{\mathrm{n}+}$ and $H_{\mathrm{n}-}$ also shift systematically with power (Fig.~5(c)). $H_{\mathrm{n}+}$ and $H_{\mathrm{n}-}$ move in opposite directions with increasing power and undergo sign reversals similar to those observed in the temperature-dependent measurements, consistent with an electrothermal change in domain-nucleation conditions. At power densities of approximately 6.2~nW/$\mathrm{\mu}$m$^2$ and above, the abrupt feature in the OOP loops is no longer readily identifiable, making $H_{\mathrm n}$ difficult to determine reliably. In addition, the OOP remanence ratio $M_{\mathrm r}/M_{\mathrm s}$ decreases sharply over the power range in which the nucleation fields cross (Fig.~5(d)), similar to its temperature-dependent evolution. Compared with the thicker Co-free device (Supplementary Fig.~S9(b--d)), the Co/FGaT device shows the OOP remanence drop and nucleation-field crossing at lower power density and has a lower OOP coercivity in this range, consistent with the Co contribution inferred from the temperature-dependent measurements. Finally, both high and low OOP remanence ratios are recovered during repeated power cycling (Fig.~5(e)), indicating that the electrothermal modulation is reversible within the tested range. Together with the complementary evolution of the OOP and IP loops in Fig.~5(a), these results support reversible switching between OOP- and IP-dominated states through Joule heating, allowing electrical control of the SRT in Co/FGaT.

\section{Discussion}
Two possible interpretations of the observed loop evolution are spin reorientation in the coupled Co/FGaT system and a change in the relative Kerr-signal weights of the two layers. In the latter case, the FGaT signal weakens near $T_{\mathrm{C}}$ while Co remains IP magnetized. A 10~nm Co film is generally expected to favor IP magnetization unless sufficient interfacial anisotropy stabilizes PMA \cite{21,33}. The heterostructure has a low-temperature OOP remanence ratio close to 1.0 (Fig.~3(c)), and each field-sweep branch exhibits one prominent jump (Fig.~3(a)). These features are consistent with interfacial exchange favoring an OOP component of the Co magnetization and coupled reversal. The layer-specific Co response remains unresolved in the present total Kerr signal.

Co/NiO/Fe trilayers provide an experimental precedent for exchange-mediated reorientation: temperature-driven in-plane spin reorientation in Fe is transferred to the Co overlayer through NiO-mediated exchange coupling \cite{24}. Here, the OOP and IP loops begin to evolve near room temperature, below the reported FGaT $T_{\mathrm{C}}$ range of approximately 350--380~K \cite{7}. The onset of this evolution is therefore consistent with anisotropy competition while FGaT remains ferromagnetic.

In addition, the Co-free device in Supplementary Fig.~S8 has the same FGaT thickness as the device in Fig.~5, providing a comparison. Over the power range in which the OOP remanence of the device in Fig.~5 decreases markedly, the Co-free device retains relatively high OOP coercivity. Its low-power coercivity reaches approximately 1~kOe (Supplementary Fig.~S8(b,d)). Another Co-free device with a thicker FGaT flake exhibits similar behavior and trends (Supplementary Fig.~S9). These observations show differences in the magnetic responses of devices with and without Co. Together with the complementary OOP and IP responses and their recovery during thermal cycling, these observations support an interpretation of the SRT in terms of temperature-dependent anisotropy competition. In this interpretation, the balance shifts between FGaT PMA and the IP preference of Co. However, the small lateral size of the device (approximately 10~$\mathrm{\mu}$m) prevented accurate, direct measurement of its temperature during external heating and electrical driving. Real-time feedback based on the device temperature was therefore impractical. We used the temperature measured near the device as a reference during external heating and power density as the input parameter during electrical driving. Considering the thermal capacitance of the substrate, possible resistance fluctuations in the AlO$_x$ heating filaments, and time-dependent thermal equilibrium, the MOKE hysteresis loops were measured within a short heating period of a few minutes. As shown in Fig. 3(a), the reference temperatures for the paired IP and OOP measurements differed by only 1--4 K. Over 294--375 K, the magnetic properties of Co/FGaT evolve monotonically and remain relatively stable against small temperature variations of 1--4 K, the minor differences between thermal cycles are not expected to affect the overall conclusions.

\begin{doublecolfigure}
  \includegraphics[width=\textwidth]{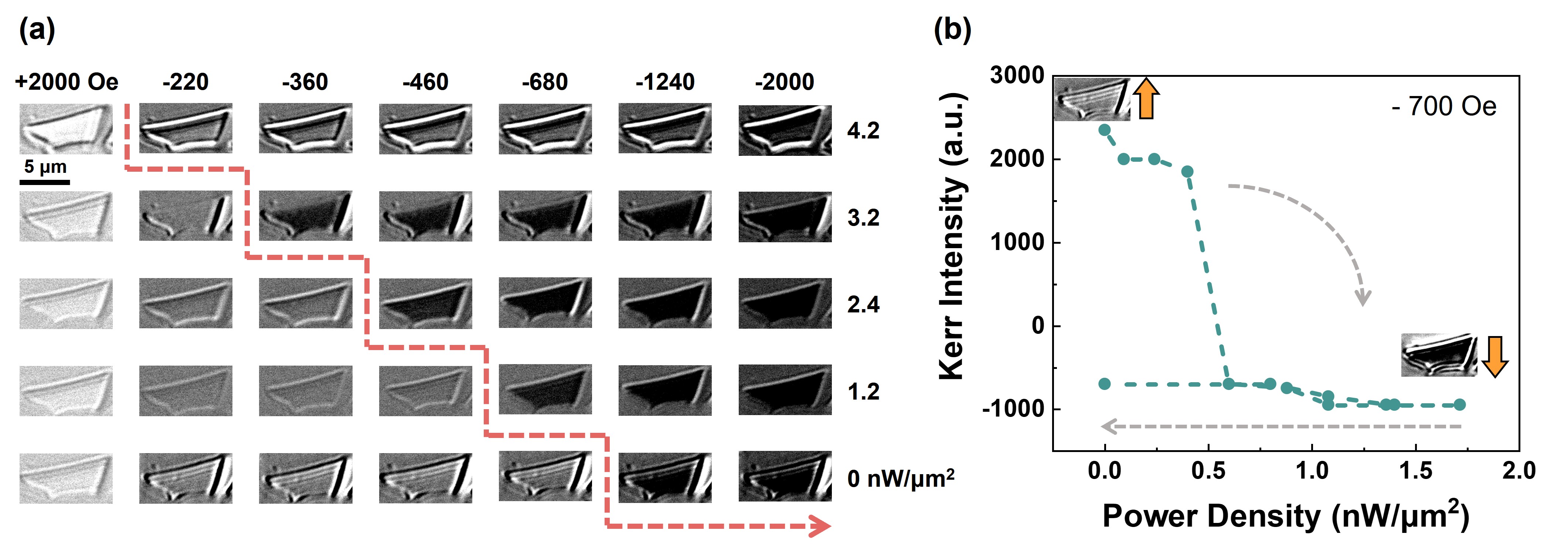}
  \caption{Power-dependent magnetization reversal in a Co-free $\mathrm{AlO}_{x}$(100~nm)/Pt(3~nm)/FGaT(39~nm) device observed by polar Kerr microscopy. (a) Polar Kerr images acquired during magnetic-field sweeps at different fixed power densities. (b) Kerr intensity as the power density is increased and then decreased under a fixed assist field of $-700$~Oe. Insets show Kerr images before and after switching.}
  \label{Domain_flip}
\end{doublecolfigure}

The polar Kerr images in Fig.~6 further show magnetization reversal in FGaT under electrical input. Measurements were performed on a Co-free $\mathrm{AlO}_{x}$(100~nm)/Pt(3~nm)/FGaT(39~nm) device. Images were recorded during field sweeps at fixed power densities between 0 and 4.2~$\mathrm{nW}/\mathrm{\mu}\mathrm{m}^{2}$ (Fig.~6(a)). As the field was swept from +2000~Oe toward negative values, the flake changed from bright to dark, indicating reversal of the OOP magnetization. Without electrical power, reversed contrast was observed at $-1240$~Oe; at 4.2~$\mathrm{nW}/\mathrm{\mu}\mathrm{m}^{2}$, it was already present at $-220$~Oe. This indicates that reversal can occur at smaller reverse-field magnitudes as the power increases. These observations support field-driven reversal facilitated by Joule heating in FGaT. The trend is consistent with the magnetic softening inferred from the temperature- and power-dependent nucleation-field shifts in Figs.~3(b) and 5(c).

Increasing the power also induced magnetization reversal under a fixed assist field (Fig.~6(b)). The applied field was held at $-700$~Oe while the power was gradually increased. The Kerr intensity dropped abruptly between 0.4 and 0.6~$\mathrm{nW}/\mathrm{\mu}\mathrm{m}^{2}$ and approached a plateau near 1.2~$\mathrm{nW}/\mathrm{\mu}\mathrm{m}^{2}$. The accompanying contrast change indicated reversal from positive to negative OOP magnetization. When the power returned to zero, the reversed state persisted under the maintained $-700$~Oe assist field. Electrical input therefore enabled power-thresholded switching at a fixed assist field. This behavior is consistent with heat-assisted magnetic writing, in which heating facilitates field-driven reversal \cite{3,4}. Such device-level electrothermal control also complements our previous work on the localized creation of bubble domains in FGaT using a conductive probe \cite{18}.

Finally, we used X-ray absorption near-edge structure (XANES) measurements to examine the local Fe environment in FGaT after thermal treatment across $T_{\mathrm{C}}$. The measurements were performed on a related PMA heterostructure, Pd(3~nm)/[Co(0.3~nm)/Pd(0.8~nm)]$_{10}$/Pd(3~nm), grown on a bulk FGaT crystal (Supplementary Fig.~S10). Fe K-edge spectra were acquired after heating the sample above the FGaT $T_{\mathrm{C}}$, followed by FC or ZFC. These cooling protocols were similar to those used for the MFM images in Figs.~1(f) and 1(g). The absorption-edge positions and near-edge line shapes nearly overlap for FC and ZFC, with no evident differences associated with Fe valence or local coordination. In contrast, the domain patterns in Figs.~1(f) and 1(g) depend on the cooling protocol. This cooling-history dependence is consistent with reports of stripe and skyrmion textures formed through different magnetic-field and temperature histories in FGaT and $\mathrm{Fe}_{3}\mathrm{GeTe}_{2}$ \cite{11,39}. Together with the recovery of the magnetic responses over repeated cycles (Figs.~3(c) and 5(e)), these observations support the interpretation of reversible SRT driven by magnetic anisotropy competition.

\section{Conclusion}
\textit{In situ} STM showed isolated Co islands at low deposition dose and more extensive coverage with stripe-like structures at higher dose. Their approximate alignment with Te lattice directions suggests that the substrate influences Co growth on the freshly cleaved surface. 
Temperature-dependent magneto-optical Kerr measurements reveal a transition from OOP- to IP-dominated magnetic hysteresis loops. Electrical driving induces a similar evolution of the magnetic hysteresis loops in an AlOx/Pt/Pd/Co/FGaT device, attributable to Joule heating possibly from conducting filaments in the oxide layer.
The Co/FGaT heterostructure exhibited reversible switching between OOP- and IP-dominated magnetic anisotropy under either external heating or electrothermal driving. The complementary MOKE loop evolution and recovery upon cycling support SRT driven by competition between FGaT PMA and the IP magnetization preference of Co. 
Furthermore, polar Kerr imaging of a FGaT device showed magnetization reversal at smaller reverse-field magnitudes as electrical power increased. At a fixed assist field of $-700$~Oe, the Kerr signal switched abruptly between 0.4 and 0.6~$\mathrm{nW}/\mathrm{\mu}\mathrm{m}^{2}$. These results demonstrate electrical control of field-assisted reversal alongside the reversible anisotropy switching observed in Co/FGaT devices, which could be potentially applied in 2D spintronics.

\section{Methods}
\noindent\textit{Crystal growth.}\\
High-quality Fe$_3$GaTe$_2$ single crystals were grown by chemical vapor transport (CVT). High-purity Fe granules, Ga lumps and Te pieces were mixed in a 3:1:2 stoichiometric ratio. The mixture was sealed in an evacuated quartz tube with iodine (2~mg/cm$^3$) as the transport agent. Materials were prepared in a glove box to minimize contamination. The sealed tube was held under a furnace temperature gradient of 760--710 $^{\circ}$C for 7--10 days, then cooled naturally to room temperature inside the furnace.

\noindent\textit{Sample and device fabrication.}\\
FGaT flakes were mechanically exfoliated from bulk crystals using 3M Scotch tape and transferred onto sapphire or $\mathrm{SiO}_{2}(285\,\mathrm{nm})$/Si substrates. The Si wafer was 500~$\mathrm{\mu}$m thick. Co and Pd layers were deposited at room temperature using a home-built electron-beam evaporation system with a base pressure of $5 \times 10^{-9}$ mbar. Pt and $\mathrm{AlO}_{x}$ layers were prepared by pulsed laser deposition using a LOTIS TII Nd:YAG laser. The laser wavelength, repetition rate and pulse width were 266~nm, 10~Hz and 16--18~ns, respectively. Deposition was performed at room temperature with a base pressure of $10^{-7}$ mbar and a target-to-substrate distance of 5~cm. Pt and $\mathrm{Al}_{2}\mathrm{O}_{3}$ targets were used for the Pt and $\mathrm{AlO}_{x}$ layers, with laser fluences of 2.8 and 1.6~J/cm$^2$, respectively.

\noindent\textit{Scanning tunneling microscopy.}\\
Surface morphology and atomic structure were characterized using a low-temperature ultrahigh-vacuum (UHV) STM (Unisoku USM-1300S with an RHK R9 controller)~\cite{40}. Images in this study were acquired at 4~K using electrochemically etched W tips. To minimize oxidation, the crystals were cleaved in a preparation chamber at $10^{-6}$ mbar and then transferred to the UHV STM stage. Co was deposited \textit{in situ} at room temperature in the same preparation chamber, after which the sample was returned to the 4~K stage for imaging. The apparent Co coverage was estimated from leveled STM topographs using Gwyddion~\cite{27}. An absolute height threshold was set separately for each image to generate a binary mask of protruding regions attributed to Co deposition. The masked area was divided by the total scan area. Threshold masks and analysis parameters, including grain-area filters, are provided in Supplementary Fig.~S1 and Table~S1.

\noindent\textit{Atomic force and magnetic force microscopy.}\\
Topography and flake thicknesses were measured using a Bruker Innova AFM with silicon tips having a resonant frequency of 300~kHz and a spring constant of 26~N/m. MFM measurements used CoCr-coated tips on the same system. The FGaT bulk crystal was heated above $T_{\mathrm{C}}$, then cooled to room temperature in zero field (ZFC) or a 440~Oe field along the c axis (FC). The 440~Oe field was applied only during FC. MFM images were acquired at room temperature under ambient conditions without an applied magnetic field.

\noindent\textit{Magneto-optical Kerr effect measurements.}\\
MOKE hysteresis loops were recorded under ambient conditions in the polar and longitudinal geometries using an Evico Magnetics GmbH microscope~\cite{41,42}. These geometries probed the OOP and IP magnetic responses, respectively. The magnetization-reversal images in Fig.~6 were acquired in the polar geometry as background-subtracted difference images referenced to the saturated state. For temperature-dependent measurements, the sample was mounted on a home-built heating stage and its temperature was monitored with a K-type thermocouple. For electrothermal measurements, a Keithley 2400 source-measure unit applied the bias and recorded the current in the two-terminal configuration described below. The voltage was held constant with the oxide in its low-resistance state. After setting the stage temperature or applying the bias, we waited a few minutes before starting the field sweep. Each MOKE loop took approximately $3\pm1$~min to record. The control parameter was the dissipated power $P=IV$ divided by the substrate area.

\noindent\textit{Device geometry and power normalization.}\\
All electrically driven devices were fabricated on 5~mm $\times$ 5~mm $\mathrm{SiO}_{2}$/Si substrates, with $\mathrm{AlO}_{x}$ covering the entire substrate. Electrothermal and current--voltage measurements used two silver-paste contacts, one on the top $\mathrm{AlO}_{x}$ surface and the other on the abraded substrate backside. Silver paste was used instead of wire bonding to avoid puncturing the thin oxide layer. Power densities were calculated by dividing the power dissipated in the entire device, $P=IV$, by the substrate area of 25~mm$^2$. This substrate-area-normalized value does not imply a uniform distribution of current or heat sources. If current is concentrated in localized conduction paths, the local dissipated power density can exceed this average. The magneto-optical measurement sampled a single FGaT flake. Its local temperature was not measured directly, and its quantitative relation to the reported power density remains undetermined.

\noindent\textit{X-ray absorption spectra.}\\
Fe K-edge XANES spectra of the related PMA heterostructure in Supplementary Fig.~S10 were acquired at Taiwan Beamline BL12B2 at SPring-8 in fluorescence mode at room temperature. Measurements were performed after heating the sample above the FGaT $T_{\mathrm{C}}$ and cooling to room temperature under FC or ZFC conditions. The XANES spectra were processed using ATHENA~\cite{43,44}, and the absorption-edge positions and near-edge line shapes were compared between the two cooling protocols.

\section{CRediT author contribution statement}
\textbf{Po-Wei Chen}: Conceptualization, Data curation, Investigation, Methodology, Writing -- original draft.
\textbf{Ming-Hsien Hsu}: Conceptualization, Data curation, Investigation, Methodology.
\textbf{Cheng-Ying Hsiao}: Data curation, Investigation.
\textbf{Ming-Yang Ho}: Data curation, Investigation.
\textbf{Masahiro Haze}: Data curation, Investigation.
\textbf{Yan-Ru Chu}: Data curation, Investigation.
\textbf{Yu-Cheng Shao}: Investigation.
\textbf{Po-Chun Chang}: Conceptualization, Methodology.
\textbf{Seng-Wui Lim}: Methodology.
\textbf{Chen-Yu Ou}: Investigation.
\textbf{Ruei Chen}: Investigation.
\textbf{Ko-Fan Chen}: Data curation, Investigation.
\textbf{Fang-Yuh Lo}: Methodology, Resources, Project administration.
\textbf{Chao-Hung Du}: Methodology, Project administration, Resources.
\textbf{Yukio Hasegawa}: Methodology, Project administration, Resources.
\textbf{Chung-Ting Ke}: Methodology, Project administration, Resources, Supervision.
\textbf{Wen-Chin Lin}: Conceptualization, Data curation, Formal analysis, Funding acquisition, Investigation, Project administration, Resources, Supervision, Validation, Visualization, Writing -- original draft, Writing -- review \& editing.

\section{Declaration of competing interest}
The authors declare no competing interests.

\section{Acknowledgments}
The authors thank the staff of the National Synchrotron Radiation Research Center (NSRRC) for technical support at Taiwan Beamline BL12B2 at SPring-8, and the members of the Hasegawa group at the Institute for Solid State Physics, the University of Tokyo, for their assistance with the STM measurements. This work was supported by ‘‘Higher Education Sprout Project'' of National Taiwan Normal University and the Ministry of Education (MOE) in Taiwan. This study is financially sponsored by the National Science and Technology Council of Taiwan under Grant Nos. NSTC 114-2112-M-003-014, NSTC 115-2112-M-003-007, and NSTC 113-2112-M-213-025-MY3.

\section{Appendix. Supplementary materials}
Supplementary material associated with this article can be found in the online version.

\section{Data availability}
The data that support the findings of this study are available from the corresponding authors upon reasonable request.

\bibliographystyle{unsrtnat}

\end{document}


\noindent
\begin{minipage}[t]{\textwidth}
  \centering
  {\fontsize{24}{28}\selectfont\bfseries \supplementTitle\par}
  \vspace{10pt}
\end{minipage}


\begin{suppfigurepage}
  \includegraphics[width=\textwidth,height=0.78\textheight,keepaspectratio]{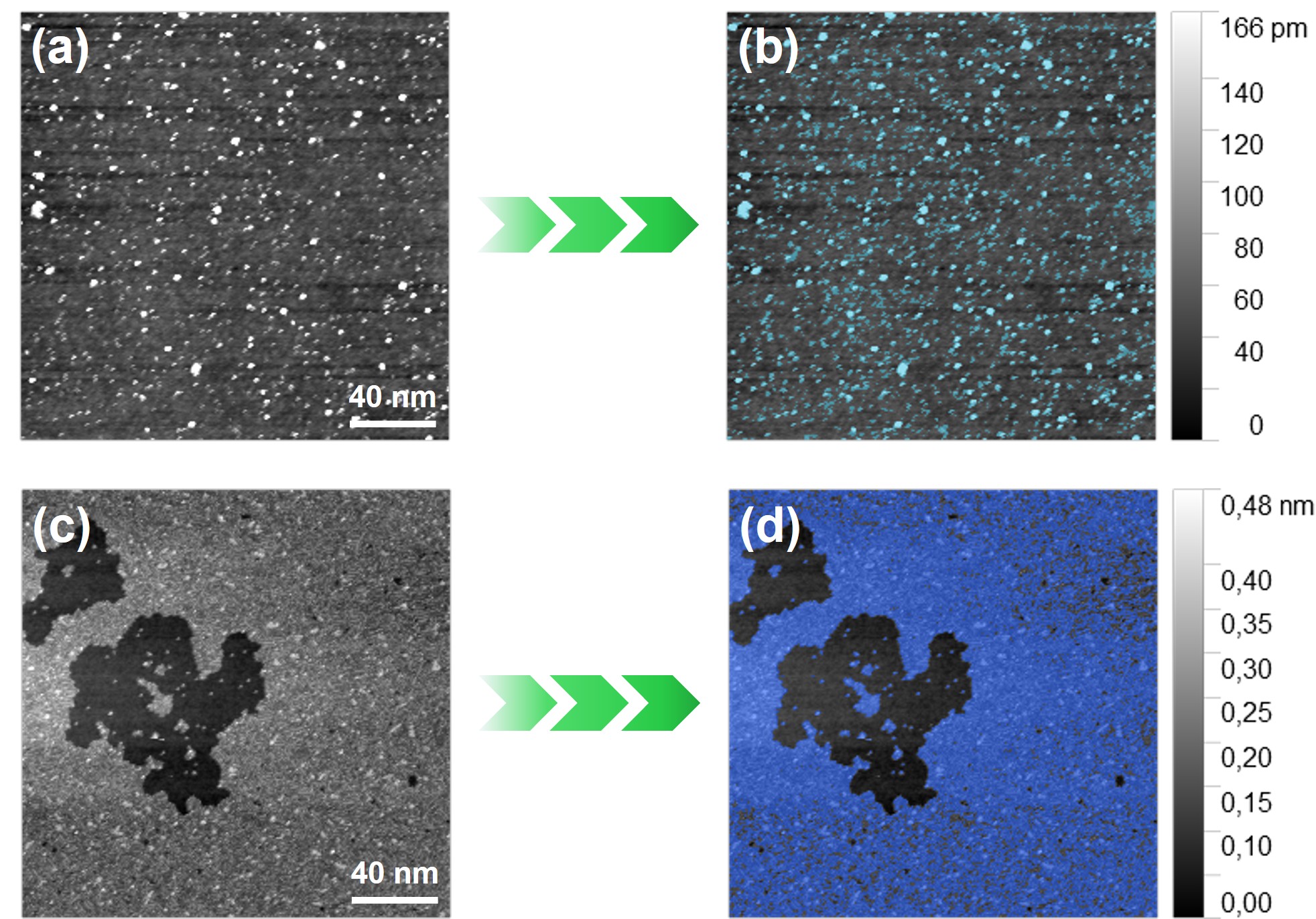}
  \captionof{figure}{Estimation of the apparent Co coverage from leveled STM topographs using Gwyddion. (a,c) Topographs acquired after lower- and higher-dose Co deposition, corresponding to Fig.~2(a,b) of the main text, respectively. (b,d) The same topographs with the corresponding threshold masks overlaid in color. The masked areas occupy approximately 13.6\% and 73.3\% of the total scan area, respectively. The height thresholds and grain-area filters used in the analysis are listed in Table~\ref{tab:co-coverage-analysis}.}
  \label{fig:s1}
\end{suppfigurepage}

\begin{center}
\captionof{table}{Parameters and results of the apparent Co coverage analysis for the two deposition doses in Fig.~\ref{fig:s1}.}
\label{tab:co-coverage-analysis}
\small
\renewcommand{\arraystretch}{1.25}
\resizebox{\textwidth}{!}{%
\begin{tabular}{lcccccc}
\toprule
Topograph &
\shortstack{Scan dimensions\\(nm $\times$ nm)} &
\shortstack{Height threshold\\(nm)} &
\shortstack{Minimum grain area\\(pixels)} &
\shortstack{Maximum grain area\\(pixels)} &
\shortstack{Masked Co area\\(nm$^2$)} &
\shortstack{Coverage\\(\%)} \\
\midrule
S1(a) & 200 $\times$ 200 & 0.06 & 2 & 72    & 5441  & 13.6 \\
S1(c) & 200 $\times$ 200 & 0.15 & 2 & 47074 & 29329 & 73.3 \\
\bottomrule
\end{tabular}%
}
\end{center}
\vspace{4pt}
\noindent
\begin{minipage}{\textwidth}
\footnotesize
\textbf{Note.}
The two rows correspond to the lower- and higher-dose topographs in Fig.~2(a,b) of the main text. A separate absolute height threshold was applied to each leveled image in Gwyddion, with the listed grain-area filters. Apparent coverage was calculated by dividing the masked Co area by the total scan area of 40,000~nm$^2$. The corresponding masks are shown in Fig.~\ref{fig:s1}(b,d).
\end{minipage}
\clearpage

\begin{suppfigurepage}
  \includegraphics[width=\textwidth,height=0.78\textheight,keepaspectratio]{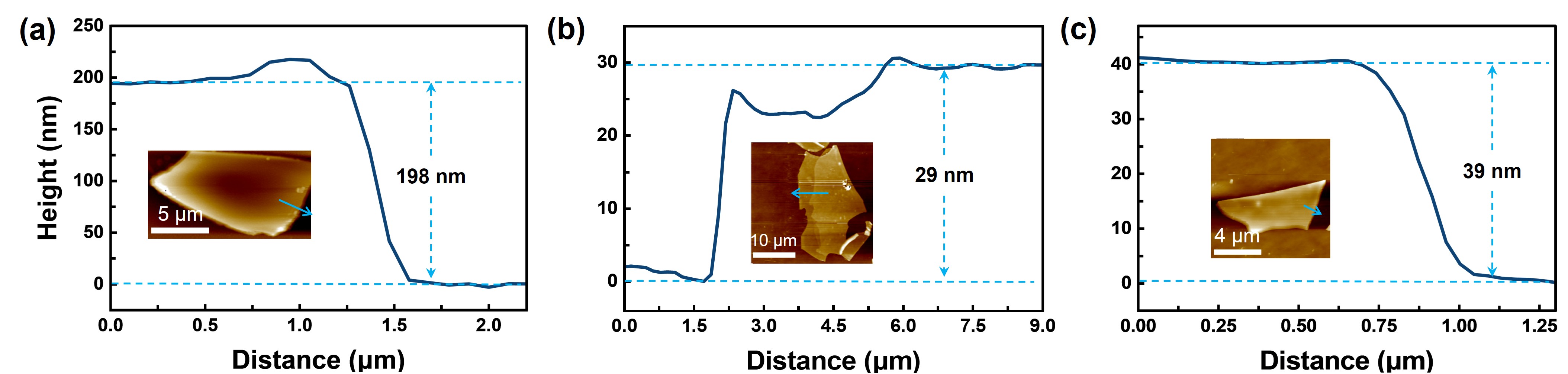}
  \captionof{figure}{AFM thickness determination of the FGaT flakes. Height profiles were taken along the arrows in the inset AFM images for (a) the 198~nm flake used in Fig.~3, (b) the 29~nm flake used in the Co/FGaT device of Fig.~5, and (c) the 39~nm flake used in the Co-free device of Fig.~6. All figure numbers refer to the main text.}
  \label{fig:s2}
\end{suppfigurepage}
\clearpage

\begin{suppfigurepage}
  \includegraphics[width=\textwidth,height=0.78\textheight,keepaspectratio]{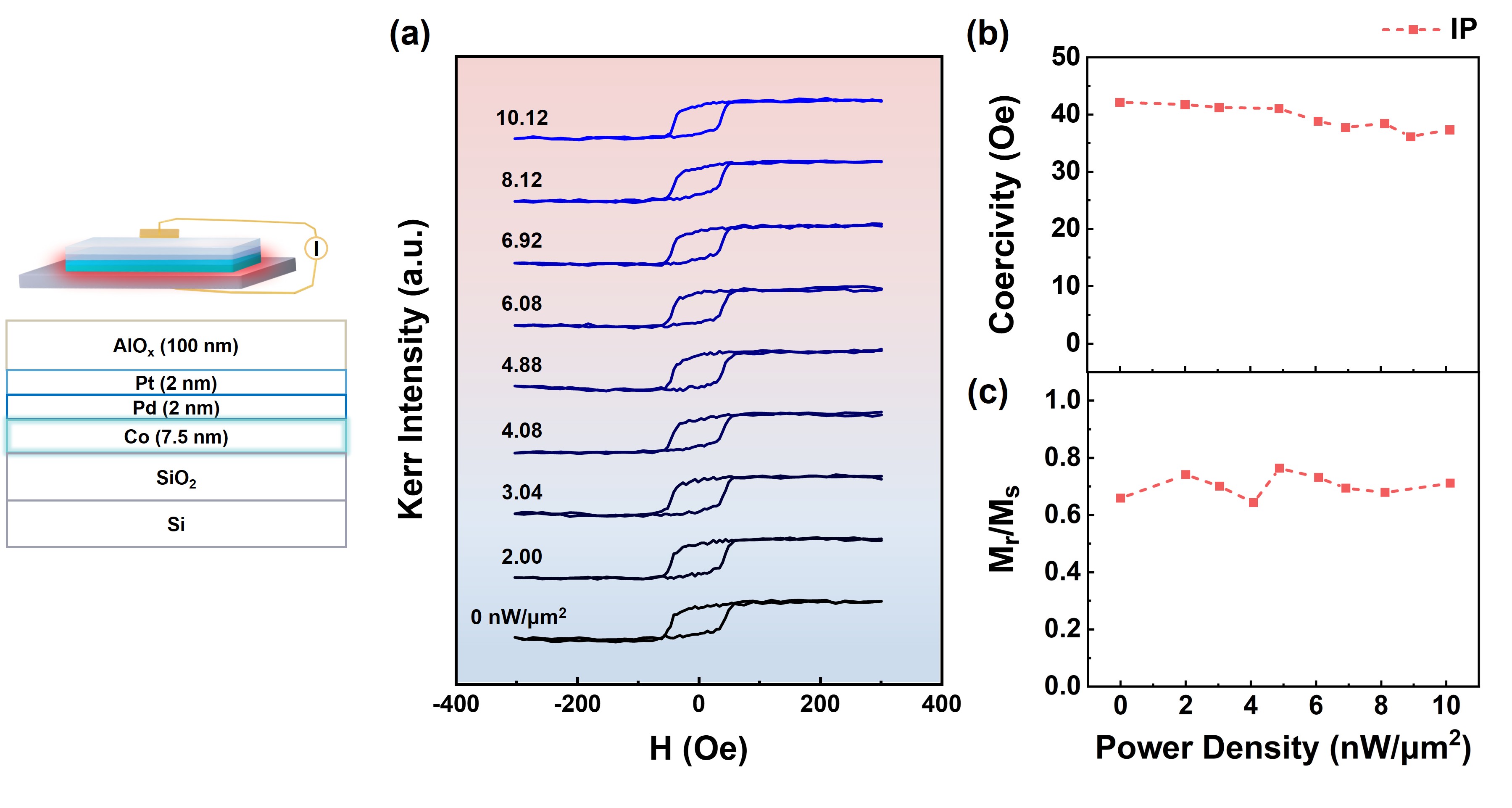}
  \captionof{figure}{In-plane magnetic response of the Co layer outside the FGaT flake in the device of Fig.~5. The layer sequence at the measurement position is AlO$_{x}$(100~nm)/Pt(2~nm)/Pd(2~nm)/Co(7.5~nm)/SiO$_{2}$/Si. (a) IP MOKE hysteresis loops at power densities from 0 to 10.12~nW/$\mu$m$^{2}$. (b) IP coercivity $H_{c}$ and (c) remanence ratio $M_{r}/M_{s}$ obtained from the loops in (a).}
  \label{fig:s3}
\end{suppfigurepage}
\clearpage

\begin{suppfigurepage}
  \includegraphics[width=\textwidth,height=0.78\textheight,keepaspectratio]{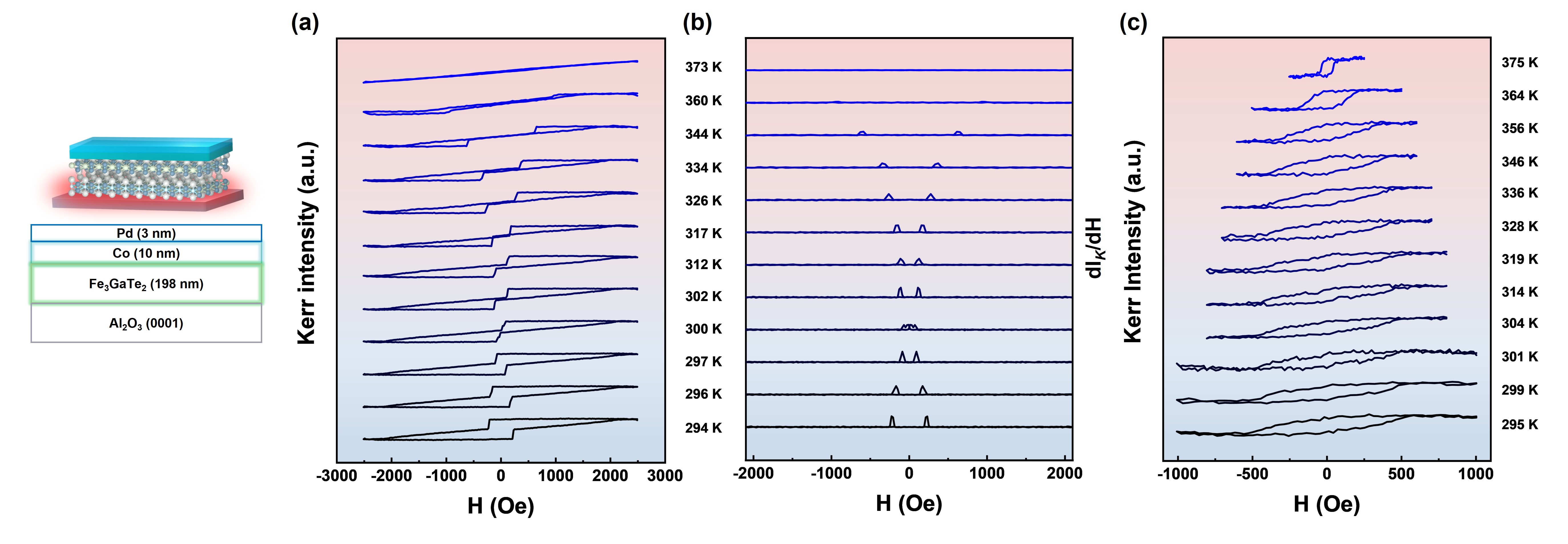}
  \captionof{figure}{Complete temperature series of the MOKE hysteresis loops and nucleation-field analysis of the Pd(3~nm)/Co(10~nm)/FGaT(198~nm) heterostructure on Al$_2$O$_3$(0001), corresponding to Fig.~3 of the main text. (a) OOP loops recorded on heating from 294 to 373~K. (b) Field derivatives of the OOP Kerr signal in (a). The characteristic peak positions on the ascending and descending field-sweep branches were used to determine $H_{\mathrm{n}+}$ and $H_{\mathrm{n}-}$, respectively. (c) IP loops recorded on heating from 295 to 375~K. Each measurement geometry includes twelve temperatures. The curves are vertically offset for clarity.}
  \label{fig:s4}
\end{suppfigurepage}
\clearpage

\begin{suppfigurepage}
  \includegraphics[width=\textwidth,height=0.78\textheight,keepaspectratio]{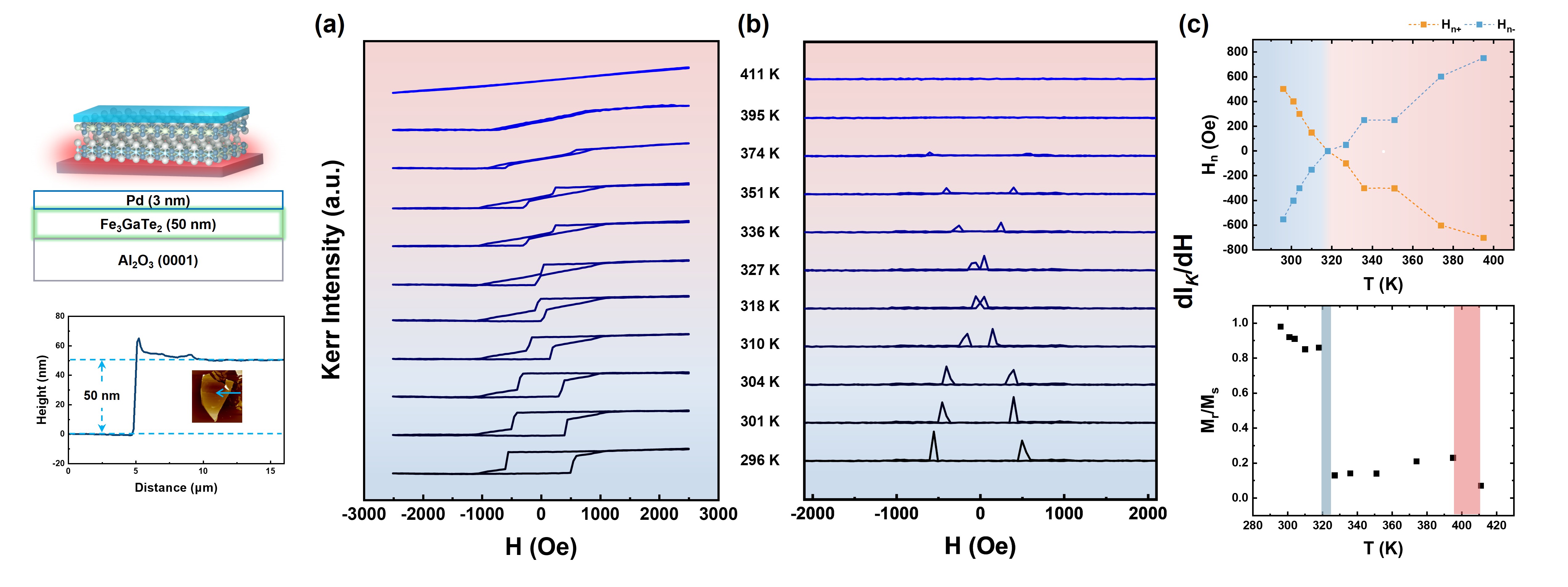}
  \captionof{figure}{Temperature dependence of magnetization reversal in a Co-free Pd(3~nm)/FGaT(50~nm) flake on Al$_{2}$O$_{3}$(0001). The AFM height profile was taken along the arrow in the inset topograph. (a) OOP MOKE hysteresis loops measured from 296 to 411~K. (b) Field derivatives of the Kerr signal in (a), shown at the corresponding temperatures. (c) Nucleation fields $H_{n+}$ and $H_{n-}$ from the ascending and descending branches (upper panel) and OOP remanence ratio $M_{r}/M_{s}$ (lower panel) as a function of temperature.}
  \label{fig:s5}
\end{suppfigurepage}
\clearpage

\begin{suppfigurepage}
  \includegraphics[width=\textwidth,height=0.78\textheight,keepaspectratio]{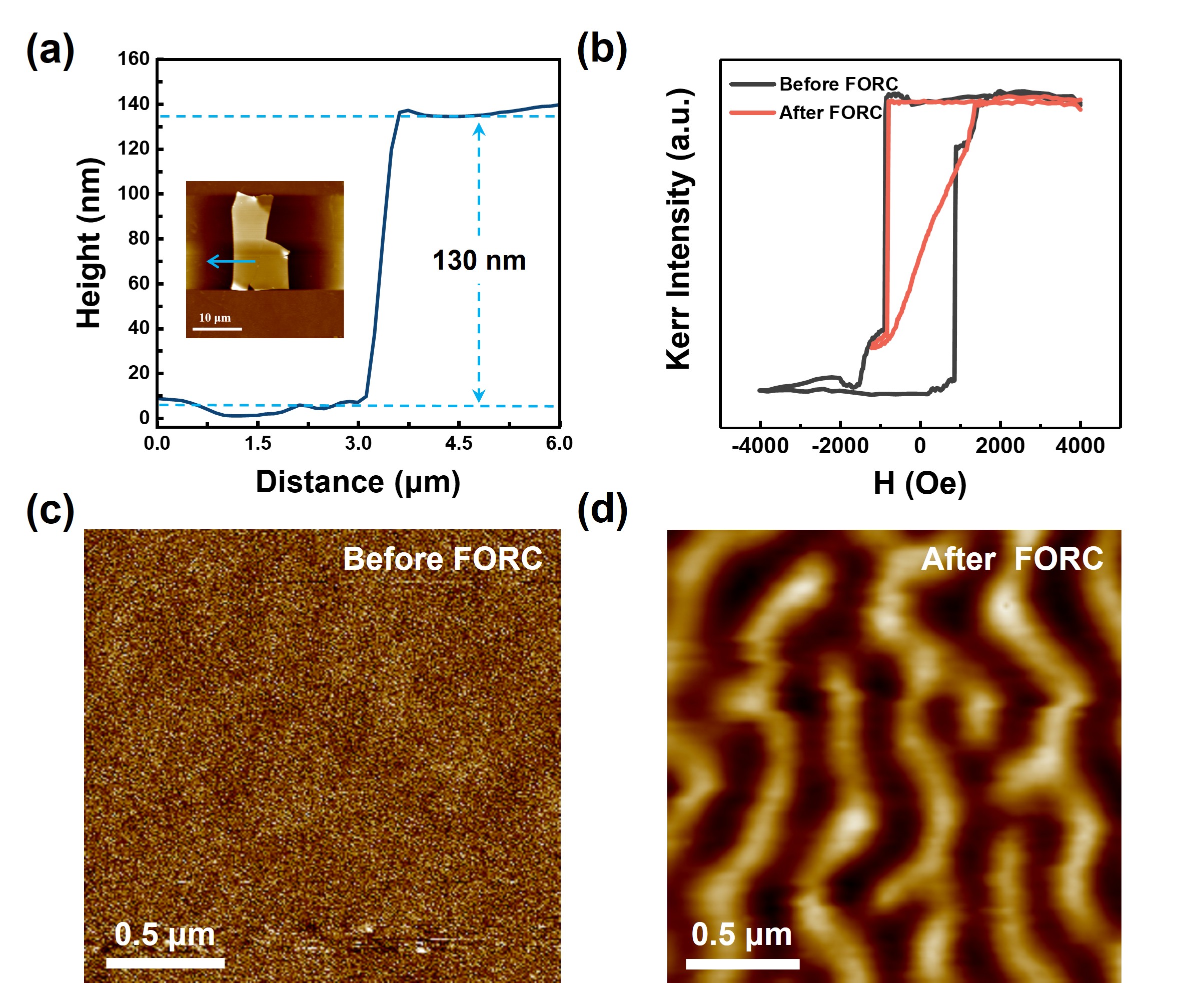}
  \captionof{figure}{Field-history dependence of the remanent domain pattern in a 130~nm FGaT flake. (a) AFM height profile along the arrow in the inset topograph. (b) OOP Kerr traces for the major loop (black) and the field excursion used to prepare the second remanent state (red). For the first-order reversal curve (FORC) protocol, the field was swept from saturation to a reversal field near $H_{n}$ and then swept back toward saturation. (c,d) MFM images at zero applied field. No domains are resolved on returning directly from saturation (c). After the excursion to the reversal field, alternating bright and dark domains are visible on returning to zero field (d). Scale bars in (c,d), 0.5~$\mu$m.}
  \label{fig:s6}
\end{suppfigurepage}
\clearpage

\begin{suppfigurepage}
  \includegraphics[width=\textwidth,height=0.78\textheight,keepaspectratio]{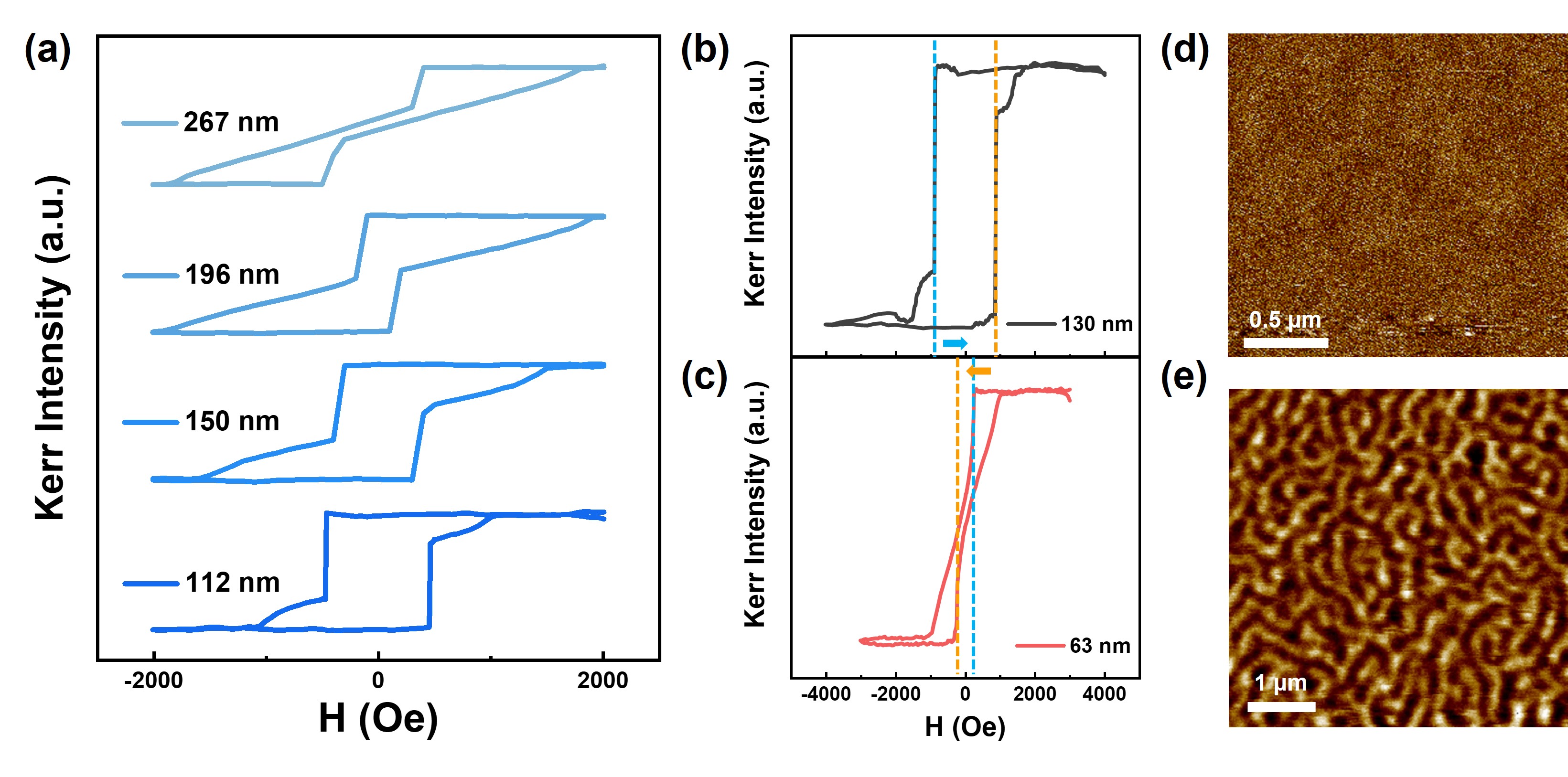}
  \captionof{figure}{Thickness dependence of magnetization reversal and remanent domains in FGaT. (a) OOP MOKE hysteresis loops for flakes of thickness 112, 150, 196 and 267~nm, vertically offset for clarity. (b,c) OOP loops for the 130 and 63~nm flakes, respectively. Dashed lines indicate the branch-specific nucleation fields. (d,e) Corresponding MFM images at remanence. No domains are resolved in the imaged region of the 130~nm flake (d), whereas the 63~nm flake shows a labyrinthine domain pattern (e).}
  \label{fig:s7}
\end{suppfigurepage}
\clearpage

\begin{suppfigurepage}
  \includegraphics[width=\textwidth,height=0.78\textheight,keepaspectratio]{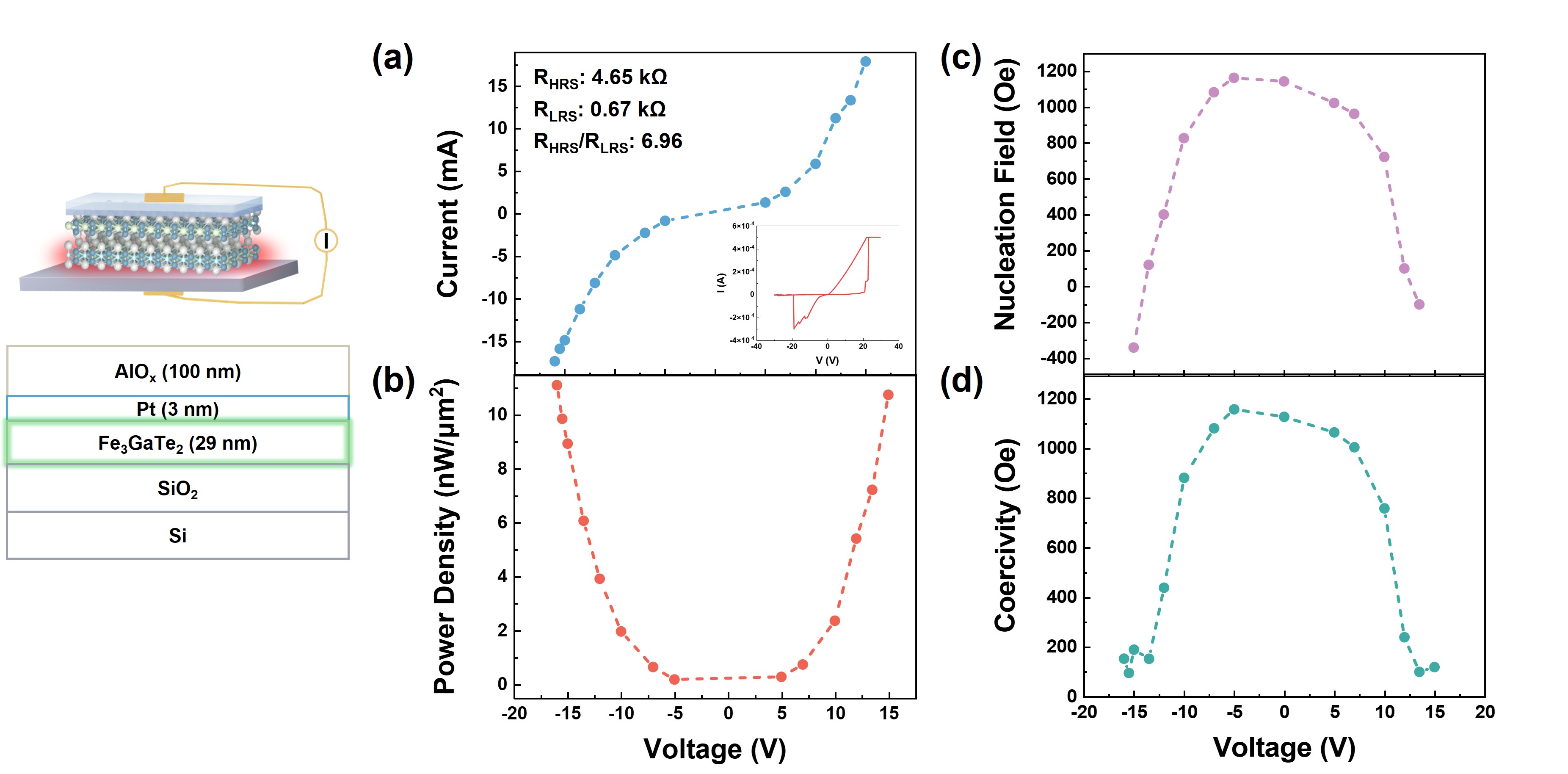}
  \captionof{figure}{Bias-voltage dependence of the electrical and magnetic properties of a Co-free AlO$_x$(100~nm)/Pt(3~nm)/FGaT(29~nm)/SiO$_2$/Si control device. (a) I--V characteristics measured across the entire device; the inset shows a hysteretic, high-resistance I--V trace. The resistance from the slope between $-5$ and $+5$~V is 4.65~k$\Omega$. The mean of the slope resistances over $+5$ to $+15$~V and $-15$ to $-5$~V is 0.67~k$\Omega$, giving a low-bias to high-bias resistance ratio of 6.96. (b) Power density, calculated as $P=IV$ divided by the substrate area of 25~mm$^2$. (c) Nucleation field $H_{\mathrm{n}}$ and (d) OOP coercivity $H_{\mathrm{c}}$ as functions of bias voltage. The similar magnetic-response trends under positive and negative bias are consistent with an electrothermal contribution. This device provides a Co-free comparison with the same FGaT thickness as the Co/FGaT device in Fig.~5 of the main text.}
  \label{fig:s8}
\end{suppfigurepage}
\clearpage

\begin{suppfigurepage}
  \includegraphics[width=\textwidth,height=0.78\textheight,keepaspectratio]{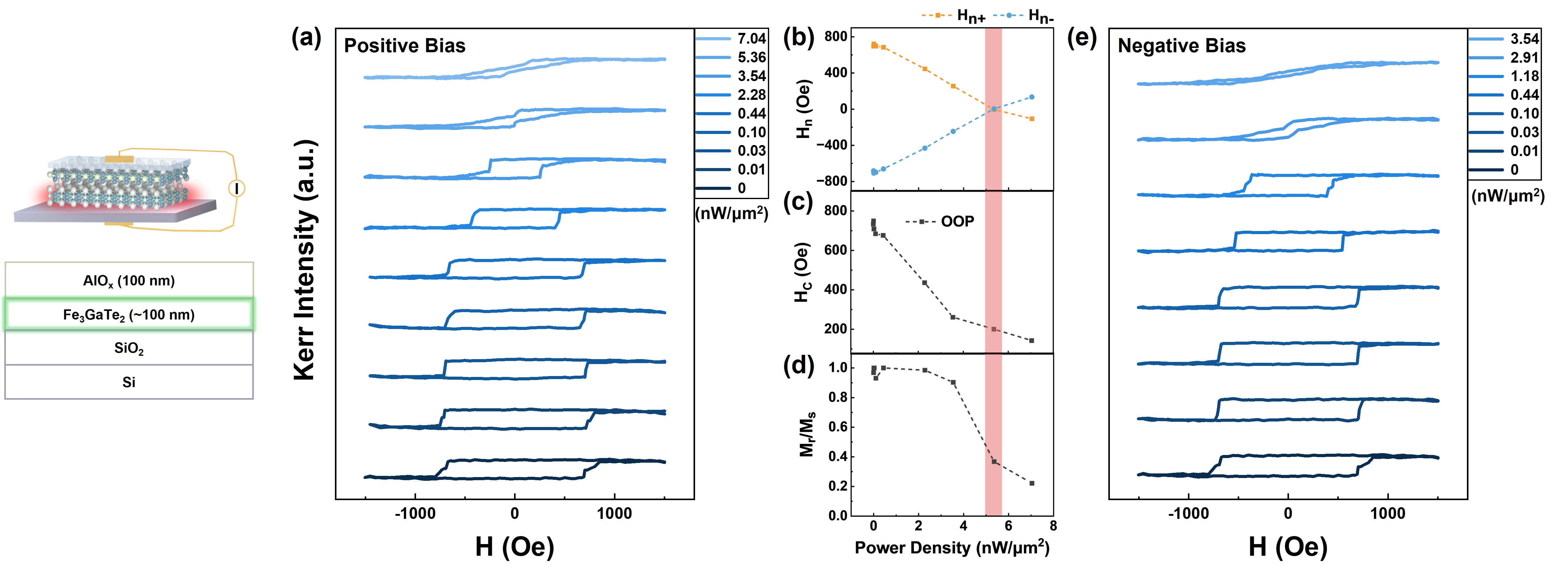}
  \captionof{figure}{Bias-polarity dependence of the electrothermal response of a Co-free AlO$_{x}$(100~nm)/Fe$_{3}$GaTe$_{2}$ (FGaT, $\sim$100~nm)/SiO$_{2}$/Si device. (a) Positive-bias and (e) negative-bias OOP MOKE hysteresis loops; the legend values denote power density in nW/$\mu$m$^{2}$, normalized by the substrate area of 25~mm$^{2}$. (b) Nucleation fields $H_{n+}$ and $H_{n-}$, (c) OOP coercivity $H_{c}$ and (d) OOP remanence ratio $M_{r}/M_{s}$ obtained from (a) as a function of power density. The red shaded band marks the power range in which the nucleation fields cross and the remanence ratio drops. The FGaT thickness was not measured directly and is estimated to be approximately 100~nm by comparing its zero-power coercivity with those of thickness-calibrated FGaT flakes. Both bias polarities produce similar qualitative loop evolution, consistent with Joule heating.}
  \label{fig:s9}
\end{suppfigurepage}
\clearpage

\begin{suppfigurepage}
  \includegraphics[width=\textwidth,height=0.78\textheight,keepaspectratio]{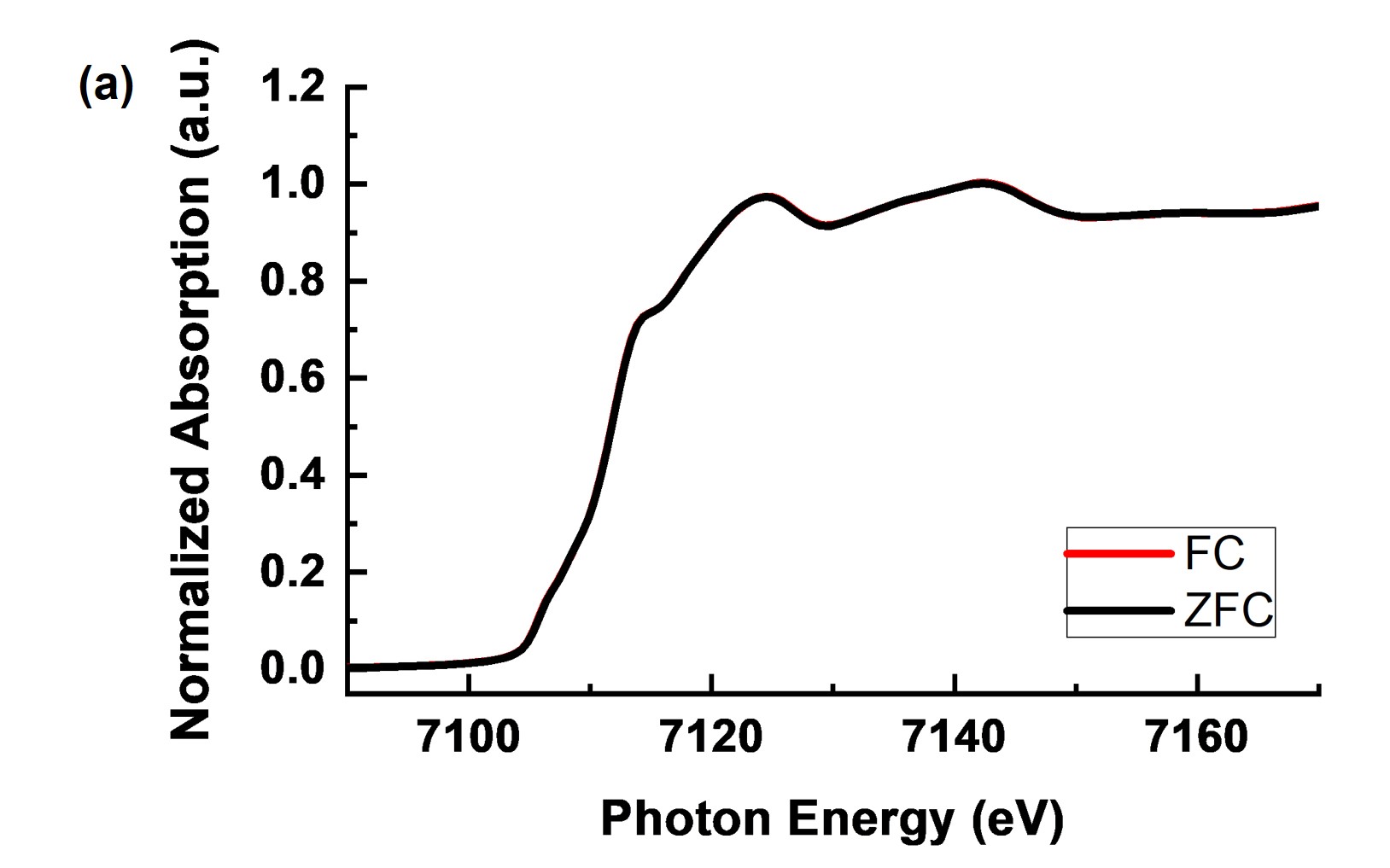}
  \captionof{figure}{Fe K-edge XANES spectra of the Pd(3~nm)/[Co(0.3~nm)/Pd(0.8~nm)]$_{10}$/Pd(3~nm) stack grown on a bulk FGaT crystal. Normalized spectra were measured at room temperature after heating above the FGaT $T_{\mathrm{C}}$ and cooling under field-cooled (FC, red) or zero-field-cooled (ZFC, black) conditions. The absorption-edge positions and near-edge line shapes nearly overlap, indicating similar local environments (local point-group symmetries) after the two cooling protocols.}
  \label{fig:s10}
\end{suppfigurepage}